\documentclass[prl,aps,amssymb,superscriptaddress,twocolumn, hypertext, showpacs]{revtex4-2} 
\usepackage{hyperref}
\hypersetup{
    colorlinks,
    linkcolor={black},
    citecolor={blue!80!black},
    urlcolor={blue!80!black}
}
\usepackage{graphicx}
\usepackage{dcolumn}
\usepackage{bm}
\usepackage{soul}
\usepackage[utf8]{inputenc}
\usepackage{amsmath}
\usepackage{multirow}
\usepackage{xcolor}
\usepackage[per-mode=symbol]{siunitx}
\usepackage{braket}
\usepackage{textcomp}

\DeclareSIUnit\gauss{G}
\graphicspath{{/}}

\makeatletter
\renewcommand*{\@fnsymbol}[1]{\ensuremath{\ifcase#1\or \dagger \or *\or \ddagger\or
   \mathsection\or \mathparagraph\or \|\or **\or \dagger\dagger
   \or \ddagger\ddagger \else\@ctrerr\fi}}
\makeatother

\begin{document}


\title{Optically coherent nitrogen-vacancy defect centers in diamond nanostructures}

\author{Laura Orphal-Kobin}
\affiliation{\vspace{0.5em}Department of Physics, Humboldt-Universit\"at zu Berlin, Newtonstr. 15, 12489 Berlin, Germany}
\author{Kilian Unterguggenberger}
\affiliation{\vspace{0.5em}Department of Physics, Humboldt-Universit\"at zu Berlin, Newtonstr. 15, 12489 Berlin, Germany}
\author{Tommaso Pregnolato}
\affiliation{\vspace{0.5em}Department of Physics, Humboldt-Universit\"at zu Berlin, Newtonstr. 15, 12489 Berlin, Germany}
\affiliation{\vspace{0.5em}Ferdinand-Braun-Institut, Gustav-Kirchhoff-Str. 4, 12489 Berlin, Germany}
\author{Natalia Kemf}
\affiliation{\vspace{0.5em}Ferdinand-Braun-Institut, Gustav-Kirchhoff-Str. 4, 12489 Berlin, Germany}
\author{Matthias Matalla}
\affiliation{\vspace{0.5em}Ferdinand-Braun-Institut, Gustav-Kirchhoff-Str. 4, 12489 Berlin, Germany}
\author{Ralph-Stephan Unger}
\affiliation{\vspace{0.5em}Ferdinand-Braun-Institut, Gustav-Kirchhoff-Str. 4, 12489 Berlin, Germany}
\author{Ina Ostermay}
\affiliation{\vspace{0.5em}Ferdinand-Braun-Institut, Gustav-Kirchhoff-Str. 4, 12489 Berlin, Germany}
\author{Gregor Pieplow}
\affiliation{\vspace{0.5em}Department of Physics, Humboldt-Universit\"at zu Berlin, Newtonstr. 15, 12489 Berlin, Germany}
\author{Tim Schröder}
\email{tim.schroeder@physik.hu-berlin.de}
\affiliation{\vspace{0.5em}Department of Physics, Humboldt-Universit\"at zu Berlin, Newtonstr. 15, 12489 Berlin, Germany}
\affiliation{\vspace{0.5em}Ferdinand-Braun-Institut, Gustav-Kirchhoff-Str. 4, 12489 Berlin, Germany}

\begin{abstract}
\noindent Optically active solid-state spin defects have the potential to become a versatile resource for quantum information processing applications. Nitrogen-vacancy defect centers (NV) in diamond act as quantum memories and can be interfaced by coherent photons as demonstrated in entanglement protocols. However, in particular in diamond nanostructures, the effect of spectral diffusion leads to optical decoherence hindering entanglement generation.
In this work, we present strategies to significantly reduce the electric noise in diamond nanostructures. We demonstrate single NVs in nanopillars exhibiting lifetime-limited linewidth on the time scale of one second and long-term spectral stability with inhomogeneous linewidth as low as 150\,MHz over three minutes. 
Excitation power and energy-dependent measurements in combination with nanoscopic Monte Carlo simulations contribute to a better understanding of the impact of bulk and surface defects on the NV's spectral properties. Finally, we propose an entanglement protocol for nanostructure-coupled NVs providing entanglement generation rates up to hundreds of kHz.
\end{abstract}

\maketitle


\section{I. Introduction}
\begin{figure*}
    \centering
    \includegraphics[width=0.90\textwidth]{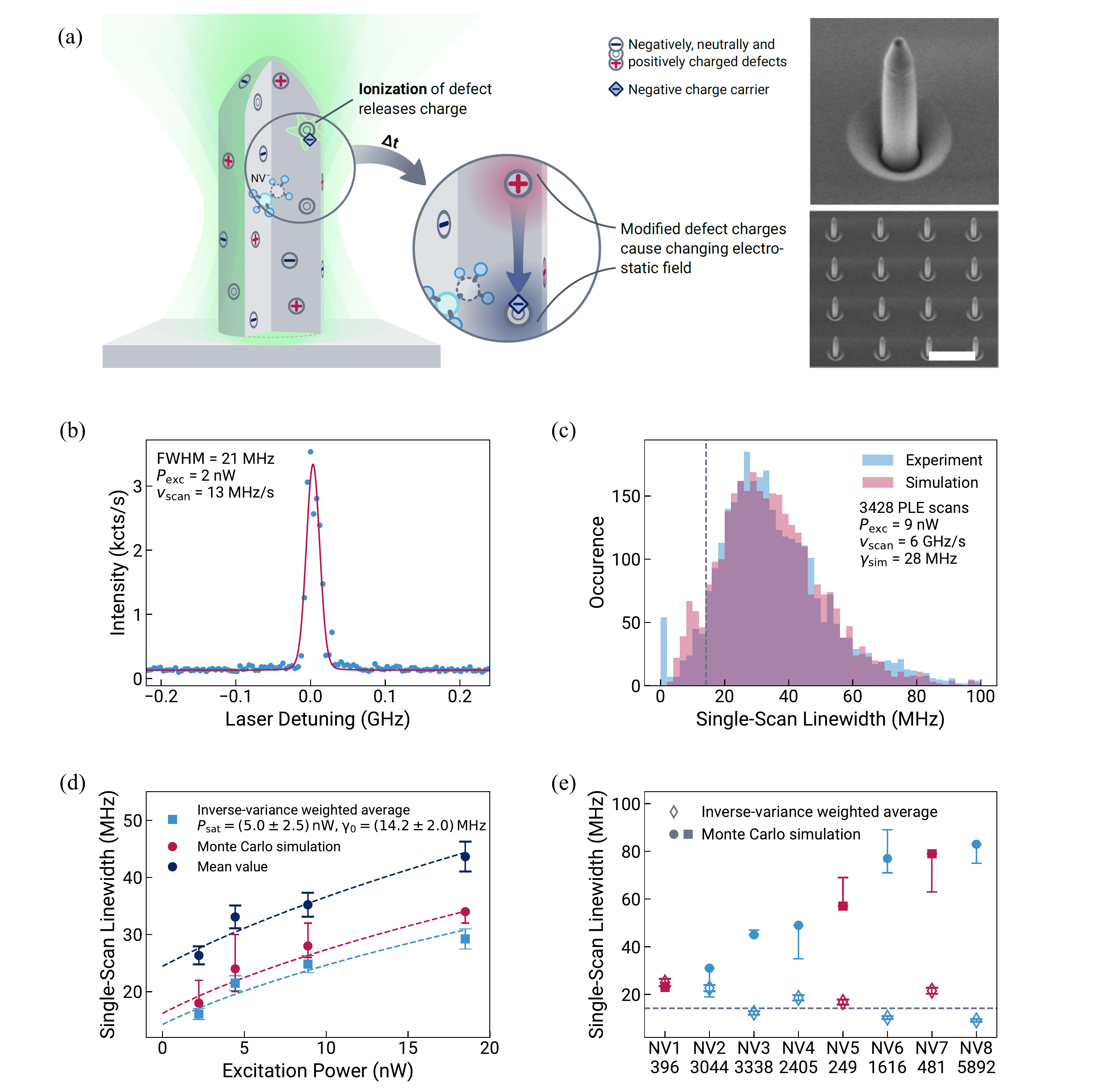}
    \caption{\label{fig:fig1} Nanostructure and single-scan linewidth. (a) Schematic sketch of an NV incorporated in a nanopillar. Green laser irradiation causes ionization of present surface and bulk defects, i.e. charges are released and move within the lattice, for example, to other defects that act as charge traps. This dynamic induces fluctuations of the electrostatic environment of the NV and leads to spectral diffusion of the NV ZPL. Insets: Scanning electron microscopy images of a single nanopillar with a radius of 125\,nm and a height of 1.6\,µm (top) and an array of identical nanopillars (bottom). The scale bar correspond to a dimension of 2\,µm. (b) A single PLE scan is described by a Voigt profile. The exemplarily selected spectrum shows a linewidth close to the lifetime-limit maintained for about one second. (c) Histogram presenting the occurrence of single-scan linewidth obtained from 3428 PLE scans. The measured data are simulated by a Monte Carlo approach. From the simulations a power-broadened linewidth of 28\,MHz is determined. The gray dashed line indicates the lifetime-limited linewidth of 14\,MHz. (d) Single-scan linewidth as a function of excitation power at a scan speed of 6\,GHz/s. The single-scan linewidths are evaluated by three methods. The data can be fitted by the function $\gamma=\gamma_0\sqrt{1+P/P_{\rm sat}}$ describing laser-induced power-broadening of the linewidth. For the Monte Carlo simulation the saturation power and natural linewidth are determined as $P_{\rm sat,MC}=(5.4\pm2.2)$\,nW and $\gamma_{\rm 0,MC}=(16.2\pm1.8)$\,MHz, and for the mean value approach as $P_{\rm sat,MV}=(8.1\pm2.9)$\,nW and $\gamma_{\rm 0,MV}=(24.4\pm2.2)$\,MHz. (e), Plot summarizing the single-scan linewidths of eight single NVs in different nanopillars from different sample regions (blue circles and red squares). The NVs are labelled with the number of PLE scans that were taken into account. The excitation power is 19\,nW and the scan speed 50\,GHz/s. In contrast to (d) we apply a 525\,nm initialization pulse before every line scan. For the linewidth analysis the Monte Carlo simulation (filled circles and squares) and inverse-variance weighting (empty diamonds) methods are used. The uncertainties of the single-scan linewidths determined by the Monte Carlo simulation are given by the $99\%$ confidence interval. For this fast scan speed the inverse-variance method is not valid anymore.}
 \end{figure*}

Optically active solid-state spin defects form a resource for a variety of quantum applications, including sensing~\cite{degen_quantum_2017}, secure communication~\cite{basset_quantum_2021,ekert_ultimate_2014}, and information processing applications~\cite{atature_material_2018,wehner_quantum_2018}. In many proposed quantum repeater protocols, quantum information transfer is achieved by interfacing spin qubits with photons~\cite{kimble_quantum_2008,borregaard_one-way_2020,kok_efficient_2005,englund_controlling_2007}. Prominently, the negatively charged nitrogen-vacancy defect center (NV) in diamond has enabled spin-photon and subsequent spin-spin entanglement of distant qubits~\cite{bernien_heralded_2013,humphreys_deterministic_2018}, recently even for a three-node quantum network~\cite{pompili_realization_2021}. However, all these impressive experiments have been performed with NVs in bulk samples using solid immersion lenses~\cite{hadden_strongly_2010} that are relatively inefficient photonic structures compared to nanostructures~\cite{schroder_quantum_2016}. Nanostructures have enabled high photon detection rates~\cite{wan_efficient_2018,neu_photonic_2014}, more efficient fibre coupling~\cite{burek_fiber-coupled_2017,torun_optimized_2021}, increased light-matter interaction~\cite{li_coherent_2015}, and allow for photonic integration~\cite{mouradian_scalable_2015}. These improvements are crucial requirements for further quantum technology development, but to date NVs in any nanostructure have not yet enabled the generation of coherent single photon states, which are the basis of any entanglement protocol. Lifetime-limited optical transition linewidths for NVs in nanostructures are hindered by the effect of spectral diffusion~\cite{rodgers_materials_2021,faraon_coupling_2012,fu_observation_2009}. Defects in the vicinity of the NV act as charge traps and electron donors that can be photoionized, inducing electromagnetic noise and a fluctuating DC Stark shift $U_{\rm Stark} = -\Vec{d} \cdot \Vec{E}(t)$ of the optical transition frequency over time as illustrated in Fig.~\ref{fig:fig1}a. 
While in bulk samples NVs with almost lifetime-limited optical linewidth of 27\,MHz have been demonstrated~\cite{chu_coherent_2014}, in nanostructures the linewidth is inhomogeneously broadened to hundreds of MHz or even several GHz~\cite{mouradian_scalable_2015,faraon_coupling_2012,wolters_measurement_2013}.
The NV charge environment depends strongly on the material properties and fabrication process, i.e. NV formation, defect density, and diamond structuring methods, as well as the distance of the NV to the surface. In recent works, it was shown that narrow linewidth NVs can be found predominantly for NVs formed from native nitrogen rather than for NVs created by ion implantation, which introduces local lattice damage such as vacancies ~\cite{van_dam_optical_2019,kasperczyk_statistically_2020}. For shallow NVs, surface defects are considered the main source of charge noise~\cite{chakravarthi_impact_2021-1}.   
Although significant efforts have been made over the past decade, the development of nanofabrication techniques~\cite{ruf_optically_2019,lekavicius_diamond_2019} and surface termination~\cite{sangtawesin_origins_2019} for improved material quality as well as active control schemes~\cite{fotso_suppressing_2015} have not yet been able to sufficiently prevent spectral diffusion in diamond nanostructures.

Here, we investigate the spectral properties of natural single NVs incorporated in diamond nanostructures with radii of 125\,nm by performing photoluminescence excitation (PLE) spectroscopy. 
We systematically examine spectral diffusion in three different excitation regimes: (1) red continuous wave (CW) illumination, (2) two-color excitation with red and higher energetic laser light, as well as (3) a shutter experiment where the NV is excited after periods of darkness. All three regimes are relevant for prospective quantum applications, for example, entanglement generation. 
We identify key parameters in control schemes for preserving spectral stability on a time scale of minutes. Furthermore, the comparison of experimental results with nanoscopic Monte Carlo simulations reveals the relation of spectral diffusion dynamics to the type and number of ionized defects. In the presented system, both bulk and surface defects contribute to an overall linewidth broadening. Finally, we propose a control protocol for efficient entanglement generation using NVs coupled to nanostructures based on the findings of this paper.

\section{II. Sample}
In contrast to most previous investigations, we select a commercial chemical vapor deposition grown sample with a relatively high density of 1\,ppm nitrogen atoms~\cite{e6_substrate_2021} resulting in a density of roughly 10\,ppb NV defect centers and a suitable yield in our NV-nanostructure device fabrication of about 10\%. Nanopillar fabrication of the $\langle 100\rangle$-oriented single-crystalline diamond substrate was carried out with dry etching processes~\cite{babinec_diamond_2010} (Appendix~A). The parameters of each step have been optimized to minimize the structural damage of the diamond surface, thus enabling repeatable high-quality fabrication over large areas as evident in the scanning electron microscopy images in Fig.~\ref{fig:fig1}a. The nanopillar radii of 125\,nm are optimized for photon coupling and the height of 1.6\,µm allows for the optical separation of nanopillar NVs from bulk NVs.

\section{III. Single-Scan Linewidth}
The linewidth of the NV zero-phonon line (ZPL) is investigated at temperatures below 4\,K by PLE spectroscopy. Here, the frequency of the excitation laser is tuned across the ZPL at 637\,nm and the photons emitted in the phonon-sideband (650-800\,nm) are detected. 
Single-scan linewidths reveal information about the indistinguishability of emitted photons. 
The data in Fig.~\ref{fig:fig1}b demonstrate lifetime-limited linewidth of a single NV embedded in a diamond nanostructure for scan speeds as slow as approximately one natural linewidth per second and for excitation powers of half of the saturation power ($P_{\rm sat}=5$\,nW, see Supplemental Material for further details~\cite{suppl_url_2022}). 
While this PLE scan is post-selected, we also statistically analyse the NV linewidth at different scan speeds and laser powers by performing thousands of PLE scans. Individual linewidths are fitted by a Voigt profile and the characteristic single-scan linewidth of a data set is determined by using the inverse-variance weighting method and a new evaluation method based on Monte Carlo simulations that we introduce in this work (Appendix~B). Due to a low number of detection events per time bin at fast scan speeds, the recorded spectra not only display a large distribution of linewidths, but produce also unphysically narrow lines below the lifetime-limited linewidth (Fig.~\ref{fig:fig1}c). This discrepancy does not allow for a direct analysis of the linewidth via simple averaging or weighting of the data, a challenge that also applies to previous works. To address this problem we employ a Monte Carlo simulation: Cauchy distributed detection events and noise are fitted with a Voigt profile. The resulting linewidths are summarized in a histogram. We then use a $\chi^2$-test to find the best linewidth corresponding to the measured data.

Excitation power-dependent measurements~(Fig.~\ref{fig:fig1}d) allow for finding a parameter regime in which the spectral stability and the negative charge state are preserved for long times while still a large number of photons are emitted. For a scan speed of 6\,GHz/s the effect of power broadening limits the preservation of lifetime-limited linewidth on average which is extracted to be ($14.2\pm 2.0$)\,MHz for zero excitation power. Comparing standard data analysis methods with our Monte Carlo approach, we find a significant offset for the mean value and only a small deviation by applying the inverse-variance weighting method. However, for faster scan speeds, linewidths determined with both previously used methods are prone to non-negligible errors (Fig.~1e). 
    	 
The observed NV properties can be similarly reproduced with NVs in other nanopillars. We investigate two sample regions containing in total 180 nanopillars. Of these, about 10$\%$ contain single NVs, i.e. appear bright in PL scans. Finally, we measure a PLE signal of eight single NVs in different nanopillars, exhibiting overall narrow linewidths with an extracted single-scan linewidth ranging from close to the lifetime-limit to about a 6-fold broadening of 83\,MHz~(Fig.~\ref{fig:fig1}e). For this overview measurement, we apply, in contrast to the other measurements, a green re-pump pulse before every line scan to ensure initialization into the negative NV charge state, leading to an additional broadening of the single-scan linewidth compared to Fig.~\ref{fig:fig1}d. All subsequent results presented in this work stem from characterization of NV3.

\begin{figure*}
    \centering
    \includegraphics[width=0.95\textwidth]{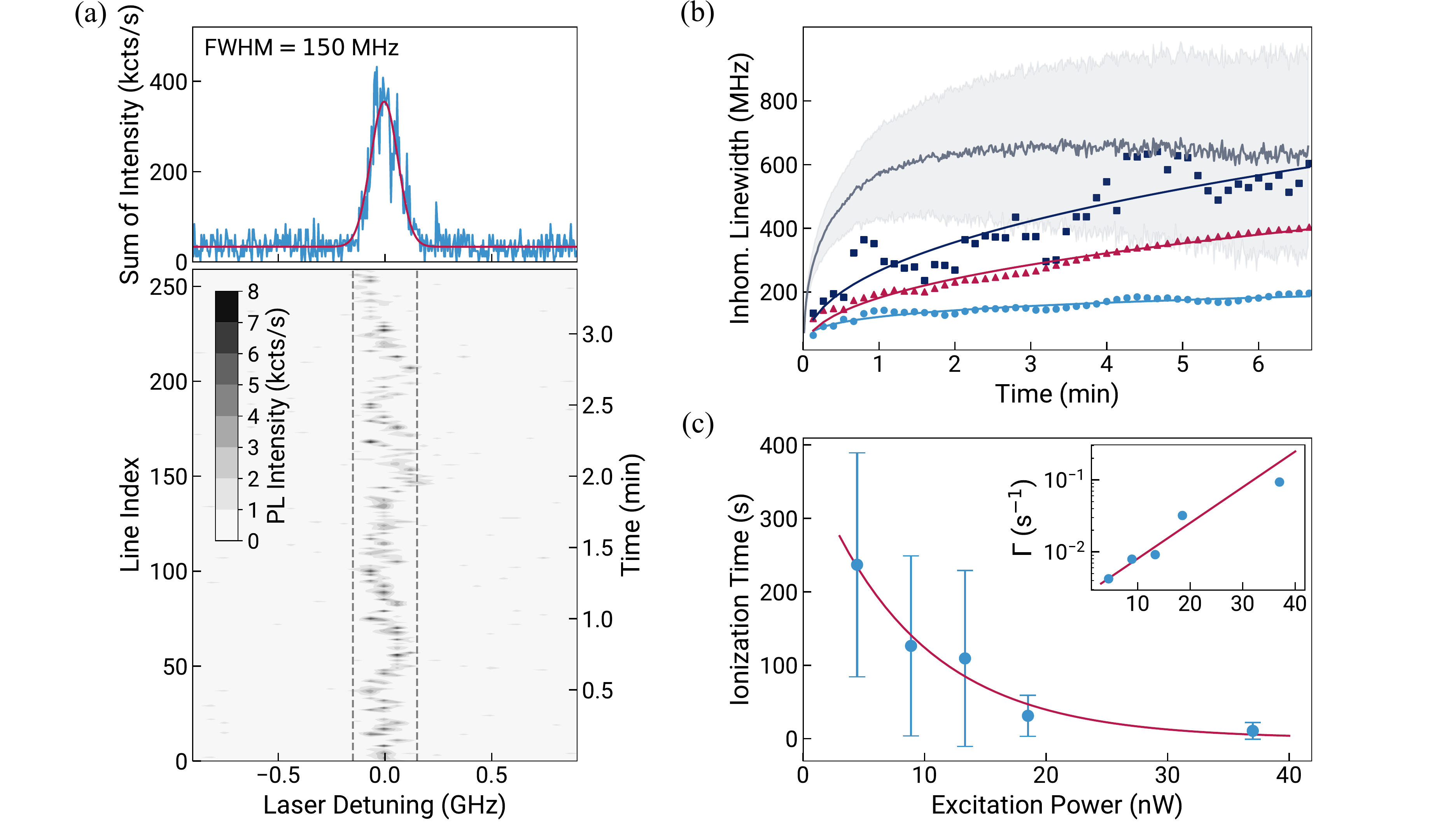}
    \caption{\label{fig:fig2} Inhomogeneous linewidth and ionization time (Regime 1). (a) More than 250 PLE scans are performed without reinitialization of the NV with green laser pulses. The excitation power of the resonant laser is 2\,nW and the scan speed is 50\,GHz/s (bottom). The inhomogeneous linewidth is given by the intensity sum of all recorded lines and is fitted by a Gaussian function. Integrated over a time of more than three minutes the FWHM is 150\,MHz (top). (b) Cumulative inhomogeneous linewidth evolution in time. Data of the single trajectory corresponding to the PLE scans in (a) (n=1, blue circles), average of post-selected trajectories (n=14, red triangles), and all trajectories (n=42, dark blue squares) are shown. The simulated data (gray line) are obtained from modelling a normal diffusion process. The experimental data are fitted by a power law $\gamma_{\rm Inh} = b P ^ a$ with exponents $a=0.22$ (blue line), $a=0.41$ (red line), and $a=0.42$ (dark blue line). A scaling exponent $a<1$ is characteristic for subdiffusive processes. (c), The average ionization time for different excitation powers is extracted from many tens of PLE data sets recorded at a scan speed of 12\,GHz/s. The ionization time decays exponentially for increased excitation power. The NV transition is driven at a duty cycle of about 0.2$\%$. Inset: Ionization rate $\Gamma$ as function of excitation power.}
\end{figure*}

\section{IV. Inhomogeneous Linewidth and Ionization Time}
Besides single-scan linewidths, which determine photon coherence on short time scales, the inhomogeneous linewidth of single NVs is an important figure of merit for the performance of quantum entanglement protocols. In this section we investigate the inhomogeneous linewidth over a time scale of several minutes. After preparing the NV in the negative charge state by using a green initialization pulse, hundreds of consecutive PLE scans are recorded at 2\,nW (half of the saturation power) without reinitialization (Regime 1). Fig.~\ref{fig:fig2}a shows an example of a range of about 250 PLE scans and the summed PL intensity recorded on a time scale of more than three minutes. 
The process of spectral diffusion, in which the transition frequency randomly shifts in time, can be clearly seen in the bottom of Fig.~\ref{fig:fig2}a. The sum of intensities of the individual scans constitutes an inhomogeneously broadened linewidth of about 150 MHz, only an 11-fold broadening compared to the natural linewidth (top of Fig.~\ref{fig:fig2}a). Although the trajectory presented here is post-selected and is not representing the average inhomogeneous linewidth, we demonstrate that under suitable fabrication and control conditions remarkably narrow linewidth can be preserved for minutes in diamond nanostructures. 
In Fig.~\ref{fig:fig2}b, the time evolution of the inhomogeneous linewidth is shown. The FWHM of the inhomogeneous line is determined from a set of trajectories summing the respective intensities for a given time of up to seven minutes, corresponding to 500 consecutive PLE scans. The data include the trajectory shown in Fig.~\ref{fig:fig2}a (blue circles) as well as the averaged values of equivalent post-selected data sets (red triangles), and the results from all measured data sets (dark blue squares). As a post-selection criterion, we reject trajectories exhibiting a spectral jump which causes a line-to-line increase of more than 200\,MHz of the cumulatively summed inhomogeneous linewidth. In 28$\%$ of the measured data, the NV is spectrally stable ($<$400 MHz) and remains in its negative charge state for seven minutes while PLE scans are performed. 
A power law fit indicates that the phenomenon of spectral diffusion induced by resonant laser irradiation is an anomalous diffusion process, more precisely a subdiffusive process~\cite{metzler_anomalous_2014}. Additionally, we use a simple stochastic process (Wiener process, Appendix~C) to model the spectral diffusion of the ZPL resonance. We also implemented the Ornstein-Uhlenbeck-model, but both approaches fail to reproduce the observed evolution of the inhomogeneous linewidth, confirming our conclusion that for the investigated system and chosen excitation parameters spectral diffusion is likely not characterized by a normal diffusion process.

In addition to spectral diffusion, charge state stability has been a challenge for NVs in nanostructures~\cite{mouradian_scalable_2015}. For investigating the ionization time, i.e. the time until the NV converts from the negative to the neutral charge state, we use again a PLE excitation scheme. The NV ZPL transition  is driven about 0.2$\%$ of the time (duty cycle). The ionization time decays exponentially as a function of excitation power (Fig.~\ref{fig:fig2}c). At excitation powers on the order of the saturation power of 5\,nW the ionization time is several minutes. During this time no green laser pulse needs to be applied for reinitializing the system, resulting in a significantly reduced spectral diffusion as evaluated in the following section. 
        
\begin{figure*}
    \centering
    \includegraphics[width=0.95\textwidth]{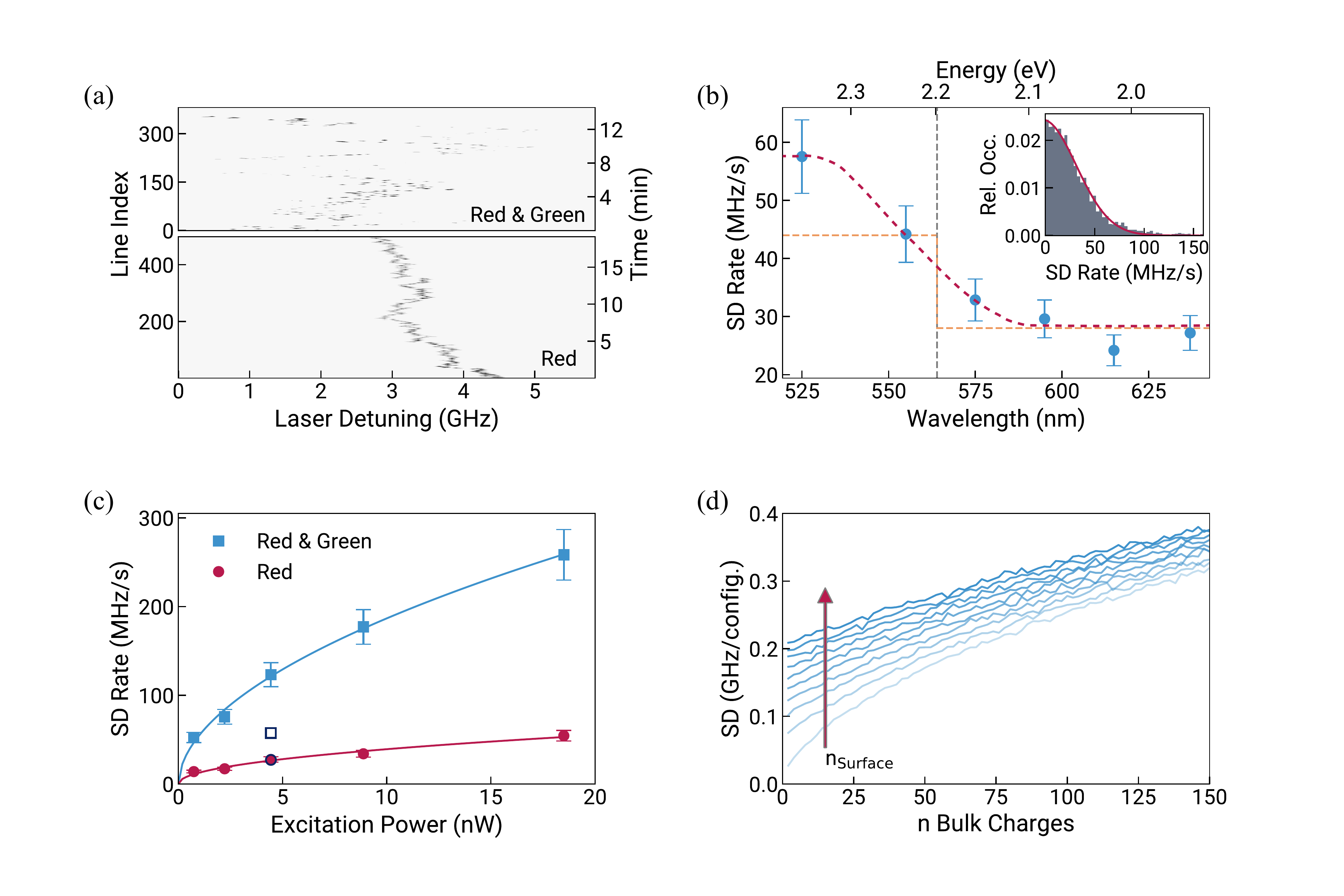}
    \caption{\label{fig:fig3} Spectral diffusion dynamics (Regime 2). (a) Hundreds of PLE scans are performed with only resonant excitation laser irradiation (bottom) and with a mixture of resonant and green (525\,nm) laser irradiation with equal ratio of a total power of 4\,nW (top). While red laser irradiation induces a small drift of the ZPL resonance frequency, additional green laser light causes large spectral jumps. (b) Spectral diffusion (SD) rate measured for different excitation wavelengths. A two-color PLE scheme, involving resonant and off-resonant laser irradiation, is applied with a total power of 4\,nW. The gray dashed line highlights the wavelength of 564\,nm which corresponds to the ionization energy of substitutional nitrogen defects. As guide for the eye the supposed spectral diffusion rate evolution is drawn, assuming bulk diamond containing only nitrogen defects (orange dashed line) and additional surface defects (red dashed line). Inset: The relative occurrence of spectral shifts can be described by a normal distribution function. Here the data excited with a mixture of resonant and 575\,nm light is shown. (c) Spectral diffusion rate as a function of the total excitation power, recorded for resonant excitation and two-color excitation with 525\,nm. The spectral diffusion rate scales like a square root power law with respect to the excitation power: $\Gamma_{\rm SDR} = b P ^ a$ with exponents $a = 0.49$ for red only and $a = 0.53$ for two-color excitation. The filled markers correspond to data recorded after ten weeks at cryogenic temperatures and the empty markers to data recorded directly after a warm-up and a second cool-down. (d) In a Monte Carlo approach, the spectral diffusion of the NV ZPL is simulated by placing a certain number of bulk and surface charges at random positions and evaluating the resulting line shift, repeated over many iterations. The number of surface charges is increased from 2 to 452 in the direction of the arrow. Different charge configurations correspond to the fluctuating charge environment during repeated PLE line scans. The increase in spectral diffusion follows an approximate square root power law by increasing the number of charges.}
\end{figure*}	

\section{V. Spectral Diffusion Dynamics}        
Although we have found comparatively long NV ionization times, occasionally the system ionizes. For the charge state conversion from NV0 to NV-, off-resonant excitation at 510 to 593\,nm~\cite{aslam_photo-induced_2013} as well as energies resonant to the NV0 transition at 575\,nm~\cite{siyushev_optically_2013} have been used, leading to different reinitialization efficiencies.
Compared to 637\,nm, the higher laser energies cause ionization of additional diamond impurities and charge dynamics which in turn lead to increased spectral diffusion, strongly reducing the optical coherence of NV defects~\cite{wolters_measurement_2013}.  
While in PLE scans with only resonant laser light (Regime 1, see above) a relatively slow drift of the ZPL resonance over time is observed, two-color excitation using a mixture of resonant and green laser light (Regime 2) causes large spectral jumps of the ZPL resonance far exceeding the natural linewidth as shown in Fig.~\ref{fig:fig3}a. This is not surprising, since the sample exhibits a high intrinsic concentration of nitrogen bulk impurities, whose ionization energies are above 637\,nm but below 525\,nm. Moreover, higher laser energies also enable activation of additional surface defects inducing further charge noise.

To date the effect of non-resonant laser irradiation on spectral diffusion of a spectrally stable NV in a nanostructure has not been investigated systematically. We therefore examine energy-dependent spectral diffusion by applying a two-color excitation scheme, using 637\,nm in combination with lower wavelength (higher energy) laser light ranging from 525 to 615\,nm, and compare it to resonant single-color PLE scans.
As characteristic quantity, we examine the spectral diffusion rate which is determined as the difference of resonance center frequencies in consecutive line scans normalized by the time step between each scan. 
In the two-color excitation experiment, we observe a significant increase of the spectral diffusion rate at wavelengths of 555\,nm (2.23\,eV) and lower (Fig.~\ref{fig:fig3}b). 
Above 555\,nm, the relatively low spectral diffusion could be caused by ionization of the relatively few lattice vacancies (GR1, 1.67\,eV)~\cite{kiflawi_electron_2007}, defect states based on hydrogen (1.2\,eV)~\cite{rosa_photoionization_1999}, or rare boron defects (0.37\,eV)~\cite{collins_optical_1993} in bulk. Below 555\,nm, the much stronger spectral diffusion is linked to the ionization of numerous nitrogen C-centers (2.20\,eV)\cite{rosa_photoionization_1999} and a variety of surface defects that are not identified individually (see below).

While energy-dependent measurements reveal the type of defect contributing to spectral diffusion, investigating the excitation power dependence provides insights into the number of defects. We find an approximate square root power law describing the increase of the spectral diffusion rate as a function of excitation power for both, one- and two-color excitation, where the magnitude of the spectral diffusion rate is significantly higher for excitation with additional green laser light (Fig.~\ref{fig:fig3}c). 
This functional relationship is qualitatively confirmed by modelling the impact of a fluctuating charge environment on the NV ZPL resonance with a Monte Carlo simulation (Fig.~\ref{fig:fig3}d, Appendix~D). For the simulation the NV is placed in the center of a cylindrical volume with the same dimensions as the investigated nanopillars and is surrounded by randomly distributed charge traps at a density of 1\,ppm according to the density of C-centers. In many iterations a fixed number of positive and negative charges are randomly distributed among the charge traps, representing the change of the charge environment during repeated line scans. A comparison of experiment and simulations indicates that there is a direct correlation between laser power and number of charges participating in the process.

Further insights into the contribution of bulk and surface defects are gained by comparing the sample after ten weeks at cryogenic temperatures and directly after a warm-up and a second cool-down. Remarkably, the overall excellent spectral properties of the investigated NV were preserved. Moreover, a comparison of the spectral diffusion rates produced by the two-color excitation at 525\,nm with a power of 4\,nW before the second cool-down with the corresponding values after the second cool-down reveals a significant reduction, whereas the spectral diffusion rate for resonant excitation remains similar (Fig.~\ref{fig:fig3}c). Although changes of the defect formation in the bulk crystal are possible, we suppose it is far more likely that the warm-up led to a release of non-ubiquitous surface defects, e.g. a layer of ice that had build up over time. Our simulations support this idea: For low excitation powers the number of participating bulk charges is relatively small. The impact of surface charges in this regime is much more pronounced and can cause a manyfold increase of the spectral diffusion rate as shown in Fig.~\ref{fig:fig3}d.

\begin{figure*}
    \centering
    \includegraphics[width=1.0\textwidth]{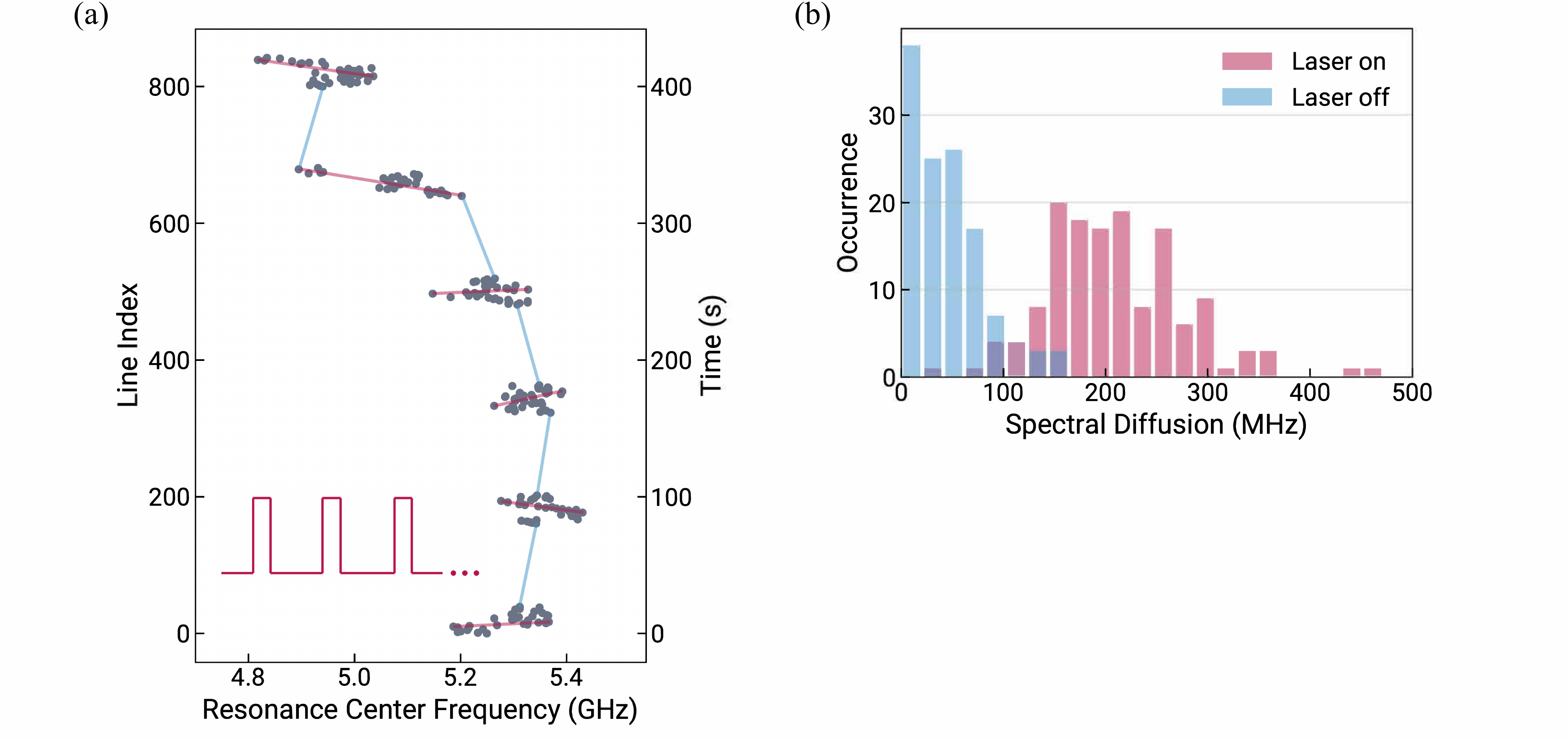}
    \caption{\label{fig:fig4} NV in the dark (Regime 3). (a) In a shutter experiment we alternate between PLE scanning for 20\,s and blocking the radiation for 60\,s. When PLE scans are performed the center frequency of the ZPL resonance is extracted from Voigt fits (gray dots). Here a data set is presented exemplarily. (b) The occurrence of spectral shifts obtained from many data sets is plotted in a histogram. The extracted spectral diffusion value for "Laser on" corresponds to the spanned frequency range recorded in the period of 20\,s. The spectral diffusion for "Laser off" is extracted from the spectral difference of the last PLE scan before and the first scan after blocking the laser as illustrated in (a).}
\end{figure*}

\section{VI. NV in the Dark}        
In previously demonstrated entanglement protocols, NVs were not exposed to laser light permanently, but resonant single-shot excitation pulses of the ZPL transitions were applied. To investigate the influence of spectral diffusion in nanostructures during darkness (Regime 3), we perform a shutter experiment.
Here, the defect is exposed to resonance-scanning laser irradiation of 2\,nW for 20 seconds and kept in darkness for 60 seconds in alternation without using green re-pump pulses~(Fig.~\ref{fig:fig4}a, Appendix~E). The frequency difference of the ZPL resonances before and after darkness, which would correspond to the difference of two single-shot read-out events, is on average 40\,MHz and the main occurrence is about 10\,MHz as summarized in the histogram in Fig.~\ref{fig:fig4}b. This measurement shows that without laser irradiation spectral diffusion is strongly reduced.

\begin{figure*}
    \centering
    \includegraphics[width=1.0\textwidth]{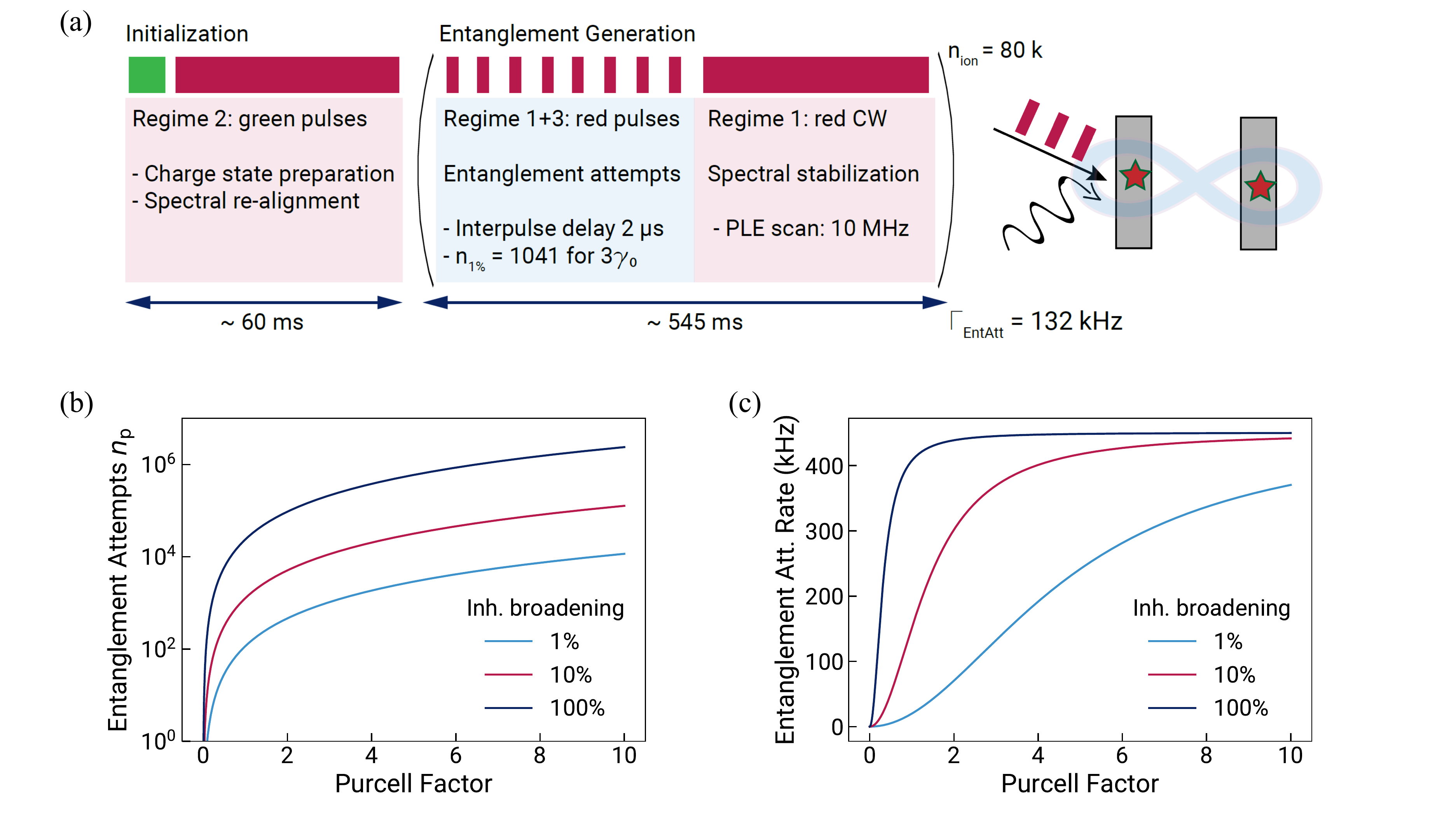}
    \caption{\label{fig:fig5} Towards entanglement generation. (a) The entanglement protocol consists of periods of system preparation (red) and entanglement generation (blue), whereas the different periods can be assigned to the three excitation regimes discussed in this work. Here, we assume a three-fold Purcell enhancement of the ZPL emission and the protocol is bounded to the condition that only entanglement attempts can be applied until the inhomogeneous linewidth is broadened by 1$\%$. Then a spectral stabilization protocol needs to be applied to tune the resonance back to the target frequency (5\,ms). When the NV ionizes the system is initialized by applying a green laser pulse. The protocol sequence for NV preparation involves charge state initialization as well as spectral re-alignment. Inset: Illustration of entanglement generation of two distant NV spin qubits embedded in nanostructures by applying coherent laser pulses (red) and microwave pulses (wavy line). (b) Entanglement attempts (number of $\pi$-pulses) that can be applied until the total linewidth broadens by $p$-percent due to spectral diffusion at a rate of 640\,MHz/s as a function of Purcell enhancement, which can be engineered in different types of nanostructures. The estimated natural linewidth of $\gamma_0=14.2\pm 2.0$\,MHz considered in the model is increased by the corresponding Purcell factor. We model the inhomogeneous broadening of the linewidth with a Wiener process. Here, an inhomogeneous linewidth broadening of $100\%$ corresponds to a broadening of twice the homogeneous linewidth. (c) Entanglement attempt rate using the protocol depicted in (a) as function of the Purcell factor. For large Purcell factors the entanglement attempt rate approaches a value of 450\,kHz which is given by the $\pi$-pulse separation of 2\,µs and the estimated ionization time of 545\,ms.}
\end{figure*}

\section{VII. Towards Entanglement Protocols}  
All previous long-distance spin-spin entanglement experiments ~\cite{bernien_heralded_2013,humphreys_deterministic_2018,pompili_realization_2021} were demonstrated with NVs embedded in solid immersion lenses, which are bulk-like microstructures that maintain optical coherence, but provide limited NV to fibre coupling efficiencies. Entanglement attempt rates as large as 182\,kHz and successful deterministic entanglement delivery at 10\,Hz were demonstrated~\cite{humphreys_deterministic_2018}. Nanostructures could potentially enhance the performance: improved design parameters enable for enhanced photon collection efficiencies~\cite{neu_photonic_2014,burek_fiber-coupled_2017, torun_optimized_2021} and increased light-matter interaction in the form of Purcell enhancement can be achieved~\cite{kaupp_purcell-enhanced_2016,li_coherent_2015, riedrich-moller_one-_2012,faraon_coupling_2012}. Considering the findings of our work, it might be possible to apply nanostructure embedded NVs for the generation of entanglement. To quantify the potential advantage, we adapt the established entanglement protocols based on $\pi$-pulse excitation to the specific requirements of our NV-nanostructure system. In particular, we analyze the protocol with respect to the three investigated excitation regimes that represent the different laser control regimes in entanglement generation, namely, (1) resonant excitation, (2) higher energetic laser irradiation for initialization, and (3) dark diffusion.

We determine resonant $\pi$-pulse parameters for the investigated NV of a duration of 2\,ns and a peak power of 3\,µW, which is technically easy to achieve (see Supplemental Material for further details~\cite{suppl_url_2022}).
From the spectral diffusion power law fit (Fig.~\ref{fig:fig3}c), we extract a spectral diffusion rate of 640\,MHz/s during excitation pulses and use this value as an input parameter in a simple stochastic model of the spectral diffusion process to evaluate the coherence of the emitted photons (Appendix~F). We assume that the inhomogeneous linewidth $\sigma_{\rm ih}$ is bounded by the diffusion law $\sigma_{\rm ih} \propto \Gamma_{\rm SDR} \sqrt{t}$, where $\Gamma_{\rm SDR}$ is the spectral diffusion rate and $t$ the evolution time. 
Based on this diffusion law, we derive the number of entanglement attempts, i.e. $\pi$-pulse excitations, that can be applied until the line is broadened by a certain amount. 
For example, an inhomogeneous broadening of 1$\%$ is a relevant regime as it still enables a Hong-Ou-Mandel visibility of around 90$\%$ and an entanglement fidelity for creating a photonic Bell-state with an optical circuit of up to 87$\%$ which we extract from ref.~\cite{kambs_limitations_2018}.

Independent of the type of nanostructure, the NV's spectral properties are assumed to be similar as presented in this work when the NV is located with at least 125\,nm distance to the surface. Considering an optical nanocavity, Purcell enhancement leads to a lifetime reduction and consequently to a broadening of the homogeneous linewidth~\cite{li_coherent_2015}. According to our stochastic model, for a fixed spectral diffusion rate a broadening of the homogeneous linewidth allows for more entanglement attempts that can be applied while optical coherence is maintained (Fig.~\ref{fig:fig5}b). Assuming a three-fold broadening of the natural linewidth, we expect to make 1041 entanglement attempts until the total linewidth is broadened by $1\%$.  Based on this result, we propose a control protocol for NVs in different types of nanostructures, taking into account the average time it takes for steady spectral re-alignment as well as system preparation after reinitialization based on PLE scans and Stark tuning~\cite{acosta_dynamic_2012} (Fig.~\ref{fig:fig5}a, Appendix~F). Entanglement attempt rates ranging from 20\,kHz up to 450\,kHz can be achieved, considering the present system with a Purcell factor set to one and other types of nanostructures that provide larger Purcell enhancement (Fig.~\ref{fig:fig5}c). Here, the maximum entanglement rate is limited by the entanglement attempt duration of 2\,µs adapted from ref.~\cite{humphreys_deterministic_2018} and system preparation time after ionization, that could both be technically improved. 

Besides reducing the sensitivity to spectral diffusion, resonant nanostructures can significantly increase the successful entanglement delivery rate by optimized fibre-coupling. In previous work, a coupling of $54\%$ of NV emission into a cavity mode was demonstrated together with a 2.7-fold Purcell enhancement~\cite{li_coherent_2015}. It was also shown that diamond cavity photons can be fiber-coupled and detected with an overall system efficiency of 85$\%$~\cite{bhaskar_experimental_2020}. Combining these achievements we expect for a cavity-coupled NV that $46\%$ of all photons are emitted into the ZPL and fibre-coupled for entanglement generation. An enhancement in ZPL emission and photon collection efficiency by more than two orders of magnitude would enable appropriate entanglement delivery rates up to hundreds of kHz.
	
 \section{VIII. Conclusion} 
 In this article, we demonstrated spectrally stable NVs in diamond nanostructures enabled by a combination of methods ranging from selection of material properties, suitable fabrication recipes, to specific control schemes. The choice of a substrate providing natural NVs and careful use of established fabrication methods facilitate reduced structural damage.
 We implemented sample specific control sequences, i.e. weak resonant excitation and reduction of high energy initialization pulses, maximally limiting charge noise. In this way, narrow inhomogeneous linewidth of the NV ZPL over minutes was demonstrated. The overall spectral stability was confirmed by recording a large number of trajectories, ensuring statistical significance of our findings. While one defect was investigated in detail, we identified seven other color centers with narrow single-scan linewidths.
 We performed a systematic characterization of the spectral diffusion of the NV ZPL resonance under different excitation regimes and investigated the contribution of bulk and surface defects by complementing the experimental results with a nanoscopic model. 
 We suppose that our methods are applicable to any other nanostructure with natural NVs located about 125\,nm apart from the surface. Therefore, a proposal of distributed NV-NV entanglement generation is presented, which can, taking advantage of increased photon collection efficiencies and Purcell enhancement in nanocavities, prospectively yield increased entanglement rates up to hundreds of kHz.
 In conclusion, we demonstrated a device based on a nanostructure-coupled NV with optical properties suitable for quantum coherent control protocols.


\appendix
\renewcommand \theequation {A\arabic{equation}} 

\section{Appendix A: Sample preparation }
The nanostructures were fabricated on a commercially available $\langle 100\rangle$-oriented single-crystalline diamond substrate, grown by chemical vapour deposition~\cite{e6_substrate_2021}. The sample surface is initially cleaned in a boiling tri-acid solution (1:1:1 of H$_2$SO$_4$ : HNO$_3$ : HClO$_4$)~\cite{brown_cleaning_2019} and subsequently etched in Cl$_2$- and O$_2$-based plasmas in order to remove any organic contaminants and structural defects~\cite{atikian_superconducting_2014}. 
The fabrication process is similar to those presented in previous works~\cite{babinec_diamond_2010}. After the deposition of a 200\,nm-thick layer of Si$_3$N$_4$ in an inductively coupled-plasma (ICP) enhanced CVD system, the sample is spin-coated with 300\,nm of electro-sensitive resist (ZEP520A) and patterned by e-beam lithography. After development, the pattern is transferred into the Si$_3$N$_4$ layer by a reactive ion etching process in a CF$_4$-based plasma (10\,sccm, RF Power\,$=$\,100\,W, P\,$=$\,1\,Pa) and subsequently etched into the diamond during an ICP process (O$_2$, 80\,sccm, ICP Power\,$=$\,750\,W, RF Power\,$=$\,200\,W, P\,$=$\,0.3\,Pa). The sample is finally cleaned in a buffered-HF solution, which completely dissolves the Si$_3$N$_4$ layer and thus exposes the diamond surface. For more details the reader is referred to the Supplemental Material~\cite{suppl_url_2022}. 

\section{Appendix B: Monte Carlo linewidth simulation}
Fast spectral scans make it difficult to determine the homogeneous linewidth of an emitter because only a few photons are collected in a single scan. Our statistical analysis shows that fitting single-line scans with a Voigt profile, which involve fewer than 20 photons, will skew the main occurrence of linewidths towards unphysically narrow lines. To ameliorate this problem, we elected to use a Monte Carlo simulation of the line scan and fitting procedure allowing us to better determine the homogeneous linewidth of spectrally unstable emitters. A single iteration of the simulation contains the following steps: A number of detection events is drawn from a Poisson distribution with a fixed mean. The events are scattered along a fixed frequency interval, weighted by a Cauchy distribution  with a fixed width $\gamma$. Additionally, a number of noise events is drawn from a Poisson distribution with a fixed mean and scattered along the frequency interval. The resulting spectrum is then fitted with a Voigt profile, and the resulting FWHM and fitting error is recorded. This is repeated for at least as many iterations as there are line scans in the experiment. We reject lines that have fewer than 5 photons in a frequency bin (4\,MHz width). Crucially, the Monte Carlo simulation allows for a change in the mean number of photon detection events ($N$) and the linewidth of the Cauchy distribution. The linewidth statistics of the Monte Carlo simulation can then be used to determine both parameters, linewidth and average photon number, by comparing them with the experimental data. As in \cite{Avni1976} we use a $\chi^2$-test to determine a best estimate of the parameters.
For more details the reader is referred to the Supplemental Material~\cite{suppl_url_2022}.  

\section{Appendix C: Stochastic diffusion model}
To determine the distribution of inhomogeneous linewidths we employ the following method: We repeatedly simulate 14 frequency trajectories whose diffusion dynamics are determined by a Wiener process. The inhomogeneous linewidth is found by fitting the trajectories at a given time step with a Voigt profile. This way we find the a distribution of inhomogeneous linewidths at a given time step. For each time step we can then calculate the mean and the variance as shown in Fig.~\ref{fig:fig2}b. 
For smaller times, the average linewidth evolution of the selected measured trajectories do not fall within one standard deviation of the simulated linewidths. This can either be attributed to the post selection or subdiffusive behaviour. For larger times the mean of the simulated trajectories deviates significantly from the square root behaviour expected for a normal diffusion process. This can be attributed to simulating only a few trajectories, which leads to large fluctuations in the fitting procedure. 
We also examined an Ornstein-Uhlenbeck process to model the diffusion of the spectral line. We could not find a good qualitative agreement between the Ornstein-Uhlenbeck process and the broadening of the spectral line for various drift parameters. For more details the reader is referred to the Supplemental Material~\cite{suppl_url_2022}.

\section{Appendix D: Monte Carlo Simulation of fluctuating electric field}
The NV's ZPL is sensitive to local electric fields due to the lack of inversion symmetry. A fluctuating charge environment produces a fluctuating local electric field at the position of the NV, which in turn Stark shifts the position of the emission line. The accumulation of these shifts over time results in an inhomogeneously broadened spectral line. 
In a single iteration of the Monte Carlo simulation we distribute a fixed number of charges in a volume that has the dimensions of the nano-pillars used in the experiment. We then determine the magnitude of the Stark shift by calculating the local electric field produced by the randomly arranged charges. For that we make use of the closed expression of the electric field produced by a point charge in a dielectric cylinder \cite{Cui_2006}. Charge screening is not systematically included in our simulations, possibly resulting in a quantitative overestimation of the accumulated local electric field fluctuations and spectral diffusion generated by a certain number of charges~\cite{oberg_solution_2020}. For more detail the reader is referred to the Supplemental Material~\cite{suppl_url_2022}.

\section{Appendix E: Shutter experiment}
The laser is switched on and off by using a mechanical shutter while PLE scans are performed continuously. Since even weak laser irradiation during the PLE scans (a few hundred ms) induces spectral diffusion, we argue that for a $\pi$-pulsed single-shot excitation on the nanosecond time scale spectral diffusion would be much more suppressed, if not fully absent.

\section{Appendix F: Entanglement protocol}
We derive a quantitative estimate of the number of entanglement attempts ($\pi$-pulse excitations) given a certain spectral diffusion rate that can be applied until the linewidth is broadened by a certain amount. We base our estimates on the Wiener process, which provides an upper bound estimation for the inhomogeneous linewidth broadening and thus a lower bound on the number of attempts.
The time dependent FWHM of a Voigt profile which broadens in time due to a random walk of the frequency is given by
\begin{equation}
    f_V(t) = a \sigma_{\rm h} + \sqrt{b \sigma_{\rm h}^2 + \sigma_{\rm ih}(t)^2} ~ ,
\end{equation}
where we used the approximation of the FWHM of a Voigt profile found in \cite{olivero_empirical_1977}. $\sigma_{\rm hom}$ is the homogeneous linewidth and the inhomogeneous contribution is linked to the spectral diffusion rate
by
\begin{equation}
\sigma_{\rm ih}(t) = \Gamma_{\rm SDR} \sqrt{4\pi \ln(2)\tau t} ~,
\end{equation}
where $\tau$ is the time step for which $\Gamma_{\rm SDR}$ was determined and $t$ is the time that has elapsed.
The two constants are  $a = 0.5346$, $b = 0.2166$.
We can calculate the time it takes to broaden the line by $p$ percent of its original value [$f_V(t_{p}) = (1+p) \sigma_{\rm h}$]:
\begin{equation}
    t_{p} = \frac{(1+p- a)^2 - b}{4 \pi \ln(2) \Gamma_{\rm SDR}^2 \tau} \sigma_{\rm h}^2~.
\end{equation}
The time $t_{p}$ is determined by the spectral diffusion rate corresponding to the $\pi$-pulse power and the homogeneous linewidth of the emitter. We assume that between the $\pi$-pulses, i.e. during darkness, spectral diffusion is not present.
Hence, in a protocol based on single-shot excitation through $\pi$-pulses with pulse duration $T_{\pi}$, the number of entanglement attempts during the time $t_{p}$ is given by 
\begin{align}
    n_{p} &= \frac{t_{\rm p}}{T_{\pi}}~.
\end{align}
The entanglement protocol consists of three parts, namely NV initialization (charge state preparation and spectral tuning), entanglement generation, and spectral re-alignment. Entanglement generation is achieved by applying resonant $\pi$-pulses with interpulse delays of 2\,µs (repetition rate of 500\,kHz) to the system. During a heralded single-photon entanglement generation attempt the spin state is prepared optically, a microwave pulse is used to create the superposition state and finally a resonant read-out $\pi$-pulse generates spin–photon entanglement followed by an additional microwave $\pi$-pulse~\cite{humphreys_deterministic_2018}.
After the estimated time during which an inhomogeneous linewidth broadening of $p$ percent is induced, a spectral re-alignment is performed. Here, the resonance is scanned in the vicinity ($\pm 10$\,MHz) of the target resonance frequency. The PLE scan is performed at a scan speed of 6\,GHz/s resulting in a scan time of about 3.3\,ms. Fitting the data and extracting the resonance center frequency would take less than 0.7\,ms. Taking time for data processing and Stark tuning into account, we estimate a spectral re-alignment time of $t_{\rm SpecCtrl}=5$\,ms when no large spectral jumps occur. The number of entanglement attempts per ionization time of 545\,ms, where entanglement generation schemes and spectral re-alignment protocols are applied alternating, is calculated by
\begin{equation}
    n_{\rm ion} = n_{p}\frac{t_{\rm ion}}{t_{p,\pi}+t_{\rm SpecCtrl}}.
\end{equation}
Here, $t_{p,\pi}$ corresponds to the entanglement attempt number $n_{p}$ multiplied with the $\pi$-pulse separation time of 2\,µs.

When the NV ionizes, it would be reinitialized by a 5\,µs green re-pump pulse and tuned again into resonance ($t_{\rm init}=60$\,ms). The high-energy initialization pulse likely causes a large spectral jump of the ZPL resonance within the inhomogeneous linewidth of 5\,GHz. To localize the ZPL resonance after the initialization pulse, a PLE scan at a scan speed of 50\,GHz/s is performed. On average it takes 50\,ms for the tunable laser to reach the resonance and induce an increase in count rate that will be detected. When the photon number passes a threshold significantly exceeding the dark count rate the scan is stopped, the resonance would be localized, and a DC voltage applied for tuning the ZPL resonance to the target frequency. For spectral fine-tuning again the spectral re-alignment scheme with PLE scans in the vicinity of the ZPL resonance frequency would be applied.
    
The final entanglement attempt rate is determined by normalizing the number of entanglement attempts during the ionization time by the total protocol duration
\begin{equation}
    \Gamma_{\rm EntAtt} = \frac{n_{\rm ion}}{t_{\rm ion}+t_{\rm init}}.
\end{equation}
For increasing Purcell factors, a saturation of the entanglement attempt rate towards 450\,kHz can be observed. The upper limit of entanglement attempt rate is given by the repetition rate of 500\,kHz as well as the ratio of entanglement generation time and system preparation time. Here, the shown entanglement attempt rate is not limited by physical laws, but technical issues that can be improved. For more details on the estimation of $\pi$-pulse parameters and the ionization time the reader is referred to the Supplemental Material~\cite{suppl_url_2022}.

\section*{Acknowledgments}
The authors would like to thank J.H.D. Munns for helpful discussions and technical support in the early stages of building the setup. L.O.-K. was supported by the state of Berlin through the Elsa-Neumann-scholarship. 
Moreover, we acknowledge funding by the Federal Ministry of Education and Research (BMBF, Project "DiNOQuant" No. 13N14921, Project QR.X with sub project No. KIS6QK4001) and the European Research Council (ERC Starting Grant “QUREP”).

\section*{Author contributions}
The confocal-microscopy setup was constructed by L.O.-K. and project specific extensions were added together with K.U. L.O.-K. carried out the PLE measurements. Programs for data analysis were developed by K.U. based on preliminary programs by L.O.-K. G.P. developed and implemented the Monte Carlo linewidth simulation, the stochastic diffusion model and the Monte Carlo simulation of fluctuating electric field. The entanglement protocol proposal was worked out by L.O.-K., G.P., and T.S. The diamond nanostructure sample was fabricated at the Ferdinand-Braun-Institute. Here, N.K. and T.P. were the diamond process responsible, M.M. performed the electron lithography, I.O. the SiN mask deposition, and R.-S.U. the dry etching process. T.S. developed the idea and supervised the project. All authors contributed to the writing of the manuscript.

\section*{Data availability}
The data that support the findings of this study are available from the corresponding author upon reasonable request.

\bibliography{bibliography}

\end{document}



\title{Supplemental Material: Optically coherent nitrogen-vacancy defect centers in diamond nanostructures}

\author{Laura Orphal-Kobin}
\affiliation{\vspace{0.5em}Department of Physics, Humboldt-Universit\"at zu Berlin, Newtonstr. 15, 12489 Berlin, Germany}
\author{Kilian Unterguggenberger}
\affiliation{\vspace{0.5em}Department of Physics, Humboldt-Universit\"at zu Berlin, Newtonstr. 15, 12489 Berlin, Germany}
\author{Tommaso Pregnolato}
\affiliation{\vspace{0.5em}Department of Physics, Humboldt-Universit\"at zu Berlin, Newtonstr. 15, 12489 Berlin, Germany}
\affiliation{\vspace{0.5em}Ferdinand-Braun-Institut, Gustav-Kirchhoff-Str. 4, 12489 Berlin, Germany}
\author{Natalia Kemf}
\affiliation{\vspace{0.5em}Ferdinand-Braun-Institut, Gustav-Kirchhoff-Str. 4, 12489 Berlin, Germany}
\author{Matthias Matalla}
\affiliation{\vspace{0.5em}Ferdinand-Braun-Institut, Gustav-Kirchhoff-Str. 4, 12489 Berlin, Germany}
\author{Ralph-Stephan Unger}
\affiliation{\vspace{0.5em}Ferdinand-Braun-Institut, Gustav-Kirchhoff-Str. 4, 12489 Berlin, Germany}
\author{Ina Ostermay}
\affiliation{\vspace{0.5em}Ferdinand-Braun-Institut, Gustav-Kirchhoff-Str. 4, 12489 Berlin, Germany}
\author{Gregor Pieplow}
\affiliation{\vspace{0.5em}Department of Physics, Humboldt-Universit\"at zu Berlin, Newtonstr. 15, 12489 Berlin, Germany}
\author{Tim Schröder}
\email{tim.schroeder@physik.hu-berlin.de}
\affiliation{\vspace{0.5em}Department of Physics, Humboldt-Universit\"at zu Berlin, Newtonstr. 15, 12489 Berlin, Germany}
\affiliation{\vspace{0.5em}Ferdinand-Braun-Institut, Gustav-Kirchhoff-Str. 4, 12489 Berlin, Germany}

\maketitle

\renewcommand \thesection {S\arabic{section}} 
\setcounter{figure}{0}
\renewcommand{\figurename}{Fig.}
\renewcommand{\thefigure}{S\arabic{figure}}
\renewcommand \thetable {S\arabic{table}} 
\renewcommand \theequation {S\arabic{equation}} 

\section*{Experiment}
\section{Sample and Fabrication}
\label{sec:A_fab}

\begin{figure}
    \centering
    \includegraphics[width=0.9\textwidth]{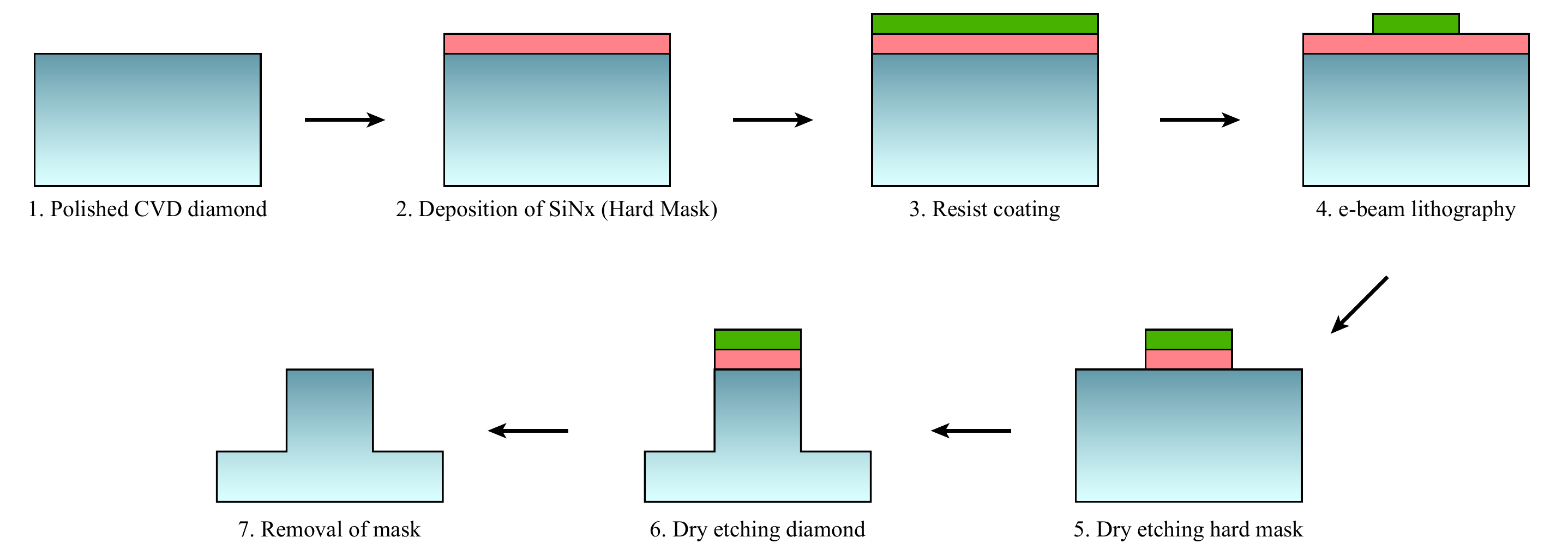}
    \caption{\label{fig:fab} Fabrication. Dry-etching fabrication process of diamond nanostructures.}
\end{figure}

The nanostructures were fabricated on a commercially available $\langle 100\rangle$-oriented single-crystalline diamond substrate, grown by chemical vapour deposition~\cite{e6_substrate_2021}. The sample surface is initially cleaned in a boiling tri-acid solution (1:1:1 of H$_2$SO$_4$ : HNO$_3$ : HClO$_4$)~\cite{brown_cleaning_2019} and subsequently etched in Cl$_2$- and O$_2$-based plasmas in order to remove any organic contaminants and structural defects~\cite{atikian_superconducting_2014}. 
The fabrication process of the nanopillars is schematically outlined in Fig.~\ref{fig:fab} and similar to those presented in previous works~\cite{babinec_diamond_2010}. After the deposition of a 200\,nm-thick layer of Si$_3$N$_4$ in an inductively coupled-plasma (ICP) enhanced CVD system, the sample is spin-coated with 300\,nm of electro-sensitive resist (ZEP520A) and patterned by e-beam lithography. After development, the pattern is transferred into the Si$_3$N$_4$ layer by a reactive ion etching process in a CF$_4$-based plasma (10\,sccm, RF Power\,$=$\,100\,W, P\,$=$\,1\,Pa) and subsequently etched into the diamond during an ICP process (O$_2$, 80\,sccm, ICP Power\,$=$\,750\,W, RF Power\,$=$\,200\,W, P\,$=$\,0.3\,Pa). We use pure O$_2$ plasma and not a mixture of O$_2$ and Ar as some groups report, which may have resulted in a reduced level of damage on the exposed diamond surface. Ar is mainly used to enhance the sputtering effect during the plasma process, usually to improve etch rate and verticality. This process, however, is quite 'physical' and energetic, as it is based on the fact that the Ar particles in the plasma are accelerated towards the exposed surfaces of the diamond and remove the material by sputtering away the C atoms. The sample is finally cleaned in a buffered-HF solution, which completely dissolves the Si$_3$N$_4$ layer and thus exposes the diamond surface.
  
A scanning electron microscopy (SEM) image of a section of the nanopillar array is shown in Fig.~1a in the main text. The nanopillars investigated in this work have a diameter of approximately 250\,nm and a height of 1.6\,µm, as estimated from SEM images. 
  
We investigated two sample regions containing 180 nanopillars. Of these, about 10$\%$ contain at least one single NV, i.e. appear bright in PL scans. In 44$\%$ of these nanopillars we were able to measure a PLE signal. No effort was put into understanding why PLE measurements were not successful in the remaining 56\%, but reasons could be fast ionization rates, or long T1 times after spin-mixing as no re-pump signal was applied.
	   
\section{Experimental Details}\label{exp. details}

\subsection*{Setup}
The nanostructures are investigated in a home-built confocal microscopy setup at cryogenic temperatures at about 3.8\,K by using a closed-cycle helium cryostat (AttoDRY800). The signal photons are collected by using an objective with a numerical aperture of 0.82 and coupling the signal into a single mode fiber. For resonant excitation a tunable diode laser (New Focus Velocity TLB-6704) is used, which is scanned by applying an external piezo voltage. Energy-dependent measurements are performed by applying a SuperK FIANIUM FIU-15 laser system with a bandwidth of 10\,nm and a repetition rate of 78\,MHz. For the wavelength read-out a wavelength meter (HighFinesse WS/7) at an integration time of 2 ms is used. Single photons are detected by two avalanche photodiodes of the series SPCM-AQRH by Excelitas Technologies. For controlling and synchronizing all devices involved in the experiment, a control software based on QuDi~\cite{binder_qudi_2017} – a modular laboratory experiment management suite – is used. All devices were interfaced with the python-based software-suite and already existing modules were adapted for own purposes. As a hardware interface an USB-6343 Multifunction I/O Device from National Instruments is used. In this work, no microwave driving or optical spin-state manipulation such as resonant excitation of the $m_s=\pm 1$ transitions is applied.

In PLE scans, we recorded single-scan linewidth while the laser was scanned back and forth. The duration of a single scan is given by the scan speed and scan range that was in most experiments about 6\,GHz. Between a back and forth scan is a break of approximately 150\,ms. To limit an error due to laser piezo hysteresis, we only consider data that was recorded when the laser is scanned in one direction for analysis of the inhomogeneous linewidth.

In all measurements, except the data presented in Fig.~1e in the main text, the NV is only initialized by a green re-pump pulse, when it was ionized after performing many individual PLE scans.

\subsection*{Two cool-downs}
The experiments were performed on a time scale of twelve weeks in total. After ten weeks, we warmed up the system to room temperature and subsequently cooled it down again to continue measurements for further two weeks. We were able to find exactly the same defect 'NV3' again we used for detailed investigations discussed in the main text. In the first cool-down, we recorded the data presented in Fig.~1b-d, 1e (red and NV3), Fig.~2, Fig.~3a and c, as well as the shutter experiment shown in Fig~4. In the second run, the data presented in Fig.~3b and 1e (blue) were recorded. By comparing the excitation power-dependent spectral diffusion rate and single scan linewidth at resonant excitation power the preservation of the overall good optical properties could be confirmed. However, for two-color excitation schemes with additional green laser light, we notice a significant difference in the spectral diffusion rate. As discussed in the main text, we assume that after a long time at cryogenic temperatures a layer of ice built up on the surface as was also already suggested in previous works. Warming and again cooling down the system lead to a release of the non-ubiquitous surface defects. Although possible, we assume that a change of the defect configuration inside the pillar is unlikely.

\section{Data Analysis}\label{sec: data analysis}

The raw PLE data are binned to a resolution of 4\,MHz. To ensure that only PLE scans providing sufficient signal photons are taken into account in the analysis, we only consider scans where at least one bin contains at least three photons.

The single-scan line profile is described by a Voigt profile that is the convolution of a Gaussian and a Lorentzian distribution function. We use a least-squares minimization algorithm for fitting, in the implementation provided by the lmfit Python package~\cite{newville_lmfit_2014}. The Voigt distribution function is modelled by
\begin{align}
    f(x;A,\mu,\sigma,\gamma) = \frac{A \textrm{Re}[w(z)]}{\sigma\sqrt{2\pi}} 
\end{align}
where 
\begin{align}
   z = \frac{x-\mu+i\gamma}{\sigma\sqrt{2}} 
\end{align}
and
\begin{align}
   w(z) = e^{-z^{2}}\textrm{erfc}(-iz) ~.
\end{align}
Here, $A$ is the amplitude, $\sigma$ corresponds to the characteristic width, and $\mu$ is the center. The FWHM is approximated by 3.6013$\sigma$.
The characteristic single-scan linewidth of a data set is determined by the inverse-variance weighted average~\cite{hartung_statistical_2008}, if the Monte Carlo method for fast scan speeds is not used (see below). The spectral diffusion rate is obtained by calculating the difference between extracted resonance center frequencies of consecutive line scans, computing the inverse-variance weighted average, and dividing it by the time between scans.

\subsection{Error Estimation and Data Processing}
The relative measurement uncertainties due to slight alignment inaccuracies is 5$\%$ for the single scan linewidth (FWHM) and 11$\%$ for the spectral diffusion rate. These values are estimated by comparing similar measurements that were performed on different days. Since the wavemeter is read out only at the beginning and end of each scan, non-liniearities due to piezo hysteresis are unaccounted for. By comparison to continuous readout, we can quantify its contribution to the relative error budget to be 6$\%$. The finite precision of the wavemeter contributes 2$\%$. Both error sources are intrinsically reflected in the overall relative measurement uncertainty.

\subsection*{Monte Carlo linewidth simulation}
Fast spectral scans, for example at scan speeds of 50\,GHz/s, make it difficult to determine the homogeneous linewidth of an emitter because only a few photons are collected in a single scan. Our statistical analysis shows that fitting single-line scans with a Voigt profile, which involve fewer than 20 photons, will skew the main occurrence of linewidths towards unphysically narrow lines. To ameliorate this problem, we elected to use a Monte Carlo simulation of the line scan and fitting procedure allowing us to better determine the homogeneous linewidth of spectrally unstable emitters. A single iteration of the simulation contains the following steps: A number of detection events is drawn from a Poisson distribution with a fixed mean. The events are scattered along a fixed frequency interval, weighted by a Cauchy distribution  with a fixed width $\gamma$. The resulting spectrum is then fitted with a Voigt profile, and the resulting FWHM and fitting error is recorded. This is repeated for at least as many iterations as there are line scans in the experiment. We reject lines that have fewer than 5 photons in a frequency bin (4\,MHz width). Crucially the Monte Carlo simulation allows for a change in the mean number of photon detection events ($N$) and the linewidth of the Cauchy distribution. The linewidth statistics of the Monte Carlo simulation can then be used to determine both parameters, linewidth and average photon number, by comparing them with the experimental data. As in \cite{Avni1976} we use a $\chi^2$-test to determine a best estimate of the parameters:

\begin{align}
    S(\gamma, N) = \sum_{i = 0}^N \frac{[O_i - E_i(\gamma, N)]^2}{E_i(\gamma, N)} ~,
\end{align}
where $O_i$ are the occurrences of the observed linewidths in a spectral bin labeled by $i$ and $E_i(\gamma, N)$ are occurrences of the simulated linewidths in the same spectral bin.  By minimizing $S(\gamma, N)$ we can determine the $\gamma$ and $N$ that best fit the observed data. Even though the simulations qualitatively reproduce a large occurrence of unphysically narrow lines they still don't match the observed data perfectly. Smaller linewidths $\gamma \leq 15 \rm\,MHz$ are therefore counted in a single bin. The binning for larger linewidths has to be adjusted according to both, the observed and simulated data because frequency bins with fewer than $5$ counts on average will make the fitting procedure more unreliable. The details of the procedure can be found in \cite{Avni1976}. The $99\%$ confidence interval for $\gamma$ can be found by determining the $\gamma$ and $N$ for which
\begin{equation}
    S(\gamma, N) \leq S(\gamma_{\rm min}, N_{\rm min}) + 9.21 ~. 
\end{equation}
The results for a range of pillars for 50\,GHz/s scan speeds and an excitation power of $19$\,nW can be seen in Fig.~1e and for a scan speed of 6\,GHz/s in Fig.~1c and d in the main text.

\subsection{Saturation Power and Power Broadening}\label{sec:psat and pb}
  
\begin{figure}
    \center
    \includegraphics[width = 1\textwidth]{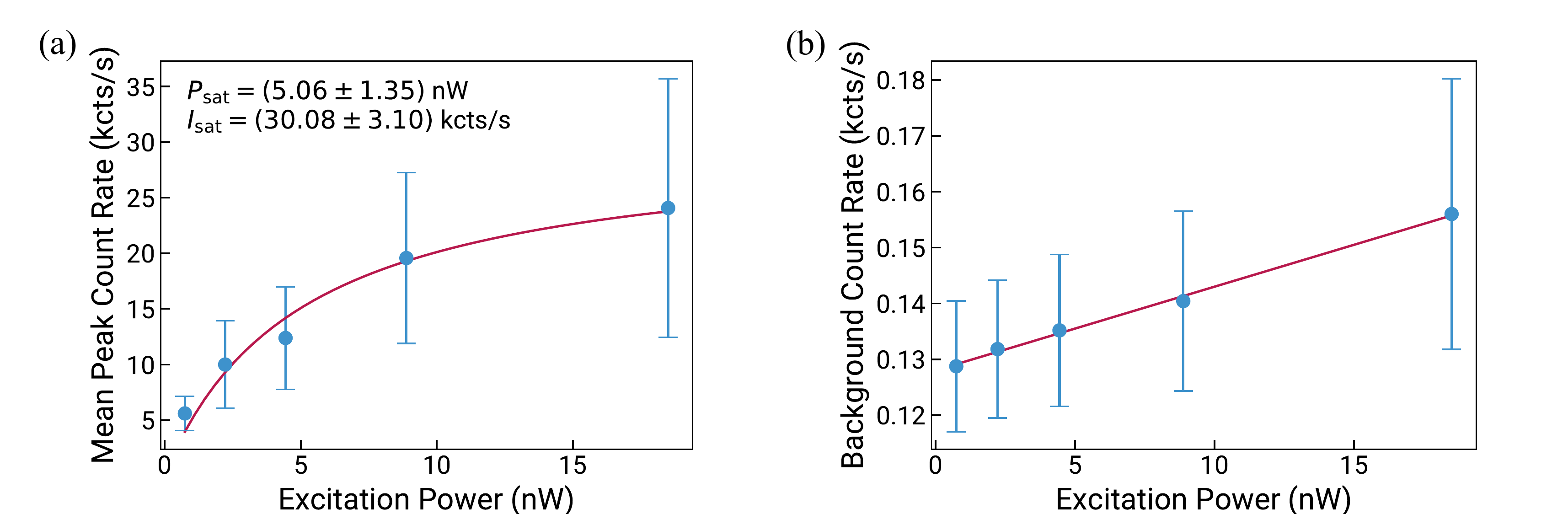}
    \caption{Saturation Measurement. (a) From the excitation power dependent PLE measurements we extract the mean peak count rate as function of the resonant excitation power (scan speed 6~GHz/s). The fit function is according to Eq.~\ref{sat}. The background at the corresponding excitation power was subtracted before. (b) The background count rate increases linearly with excitation power. The intercept at zero excitation power is at 128\,cts/s.}
    \label{Fig:sat}
\end{figure}

The saturation power of a two-level system is a characteristic parameter which, for example, will later be used to determine $\pi$-pulse parameters.
We examine two methods to extract the saturation power of 'NV3'. First, in an excitation power dependent PLE measurement the mean peak count rate is evaluated (Fig.~\ref{Fig:sat}, same data set as used for Fig.~1d in the main text). The detected intensity $I$ can be described as a function of the incident laser power $P$
\begin{align}\label{sat}
    I = I_{\rm sat}\frac{P}{P+P_{\rm sat}}~,
\end{align}
where $P_{\rm sat}$ is the saturation power and $I_{\rm sat}$ is the corresponding count rate. Before the fit is performed, we subtract the background from the signal data. From the saturation curve fit, we determine a saturation power of $P_{\rm sat}= (5.1\pm 1.4)$\,nW and a saturation intensity of $I_{\rm sat}=(30\pm 3)$\,kcts/s. 

Second, we investigate the broadening of the ZPL linewidth as function of the PLE laser power. In an ideal two-level system, for example an isolated atom, the absorption linewidth broadens with excitation power~\cite{noauthor_atomphoton_1998}
\begin{align} \label{powerbr}
    \gamma = 
                    \gamma_0\sqrt{1+\frac{P}{P_{\rm sat}}}~.
\end{align}
Here, $\gamma$ corresponds to the broadened linewidth and $\gamma_0$ to the natural linewidth at zero laser power. The expression in Eq.~(\ref{powerbr}) is used for fitting the data presented in the main text in Fig.~1d. From the power-broadening fit we obtain a saturation power $P_{\rm sat}=(5.0\pm 2.5)$\,nW and a natural linewidth of $\gamma_0=(14.2\pm 2.0)$\,MHz at zero laser power. The values we obtain for the saturation powers by applying the two different methods are in very good agreement with each other. While the obtained value for the natural linewidth indicates a Purcell enhancement in comparison to the expected bulk value of 12\,MHz, in previous work for comparable nanostructures an NV lifetime of ($14.6\pm 1.9$)~ns was observed, leading to a natural linewidth ranging from 9.6 to 12.5\,MHz~\cite{babinec_diamond_2010}. However, even though the compared structures are not identical and the position of the NVs within the nanopillars is not known, the natural linewidths are overlapping within the given uncertainties.   

To prove that the linewidths obtained from single-line scans are predominantly homogeneously broadened and the description of the broadening with Eq.~\ref{powerbr} is valid, we evaluate the spectral diffusion rate. Considering a scan speed of 5.88\,GHz/s and an average single scan linewidth of 29\,MHz at 19\,nW excitation power (compare with Fig.~1d in the main text) leads to an average time of 4.9\,ms where the excitation laser is on the resonance. The corresponding spectral diffusion rate is 49\,MHz/s. Finally, during the time of 4.9\,ms we probe the resonance with the excitation laser, the induced spectral diffusion is estimated to be 0.24\,MHz, approximately 1$\%$ of the recorded average linewidth and thus negligible.

\section{Towards Entanglement Generation \label{section:ent-attempts}}

In previous works, NVs in solid-immersion lenses were used for demonstrating remote entanglement based on resonant single-shot excitation of the ZPL transitions through $\pi$-pulses~\cite{bernien_heralded_2013,humphreys_deterministic_2018}. There, in optimized protocols an entanglement attempt duration of 5.5~µs is demonstrated corresponding to a rate of 182\,kHz, whereas a duration of less than 2\,µs ($>$500~kHz) is proposed for minor modifications of the classical experimental control protocol. As a final result a rate of 10~Hz for successful deterministic entanglement delivery was achieved. Regarding successful entanglement generation using NVs in nanostructures, we propose that for optimized protocols and system efficiencies, the lower spectral stability in comparison to NVs in solid-immersion lenses can be compensated. For NVs coupled to nanostructures, for example optical nanocavities, we expect higher emission into the ZPL and increased photon collection efficiencies, resulting in an overall increased rate of entanglement generation by orders of magnitude.

In the main text, we derive an entanglement protocol and a figure of merit for entanglement generation in diamond nanostructures taking our findings on spectral diffusion dynamics into account. From experimental results central quantities as $\pi$-pulse parameters and ionization time are estimated.  

\subsection*{$\pi$-Pulse Parameters}

From the steady-state solution of the optical Bloch equations we derive
\begin{align}\label{pisat}
    \frac{P}{P_{\rm sat}} = \frac{2\Omega^2}{\gamma_0^2}~.
\end{align}
Here, $\Omega$ is the Rabi frequency of the two-level system. 
We determine the power needed to perform a $\pi$-pulse rotation, an electric field pulse driving the ZPL transition with $\Omega T_{\pi} = \pi$, depending on the emitter lifetime $\tau_{\rm l}$, the pulse duration $T_{\pi}$, and the saturation power $P_{\rm sat}$:
\begin{align} \label{pipulse}
    P_{\pi} = 2\left(\pi\frac{\tau_l}{T_{\pi}} \right)^2P_{\rm sat}~.
\end{align}
We estimate for the investigated NV from the homogeneous linewidth a lifetime of $\tau_{\rm l} = 11.2$\,ns and use the saturation power of $P_{\rm sat}= 5.0$\,nW as parameters for a rectangular $\pi$-pulse with $T_{\pi} =2$\,ns pulse duration a corresponding power of $P_{\pi}=3.1$\,µW. 
	   
\subsection*{Ionization Time}

To get a rough estimate on the ionization time, we use the experimental data extracted from PLE measurements, presented in the main text in Fig.~2c. Here, the NV ZPL transition is only driven when the laser is on resonance. However, most of the time during the PLE scans are performed the laser is off-resonant. The duty cycle at which the ZPL transition is excited is about 0.2$\%$. 
When $\pi$-pulses of a duration of 2\,ns and a power of 3.1\,µW are applied at a repetition rate of 500\,kHz the resulting average excitation power is 3.1\,nW. 
By multiplying the experimentally determined ionization time $t_{\rm ion,PLE}=272.7$\,s (Fig.~2c in the main text) at an excitation power of 3.1\,nW with the duty cycle of 0.2$\%$ a CW ionization time of $t_{\rm ion,CW}=545.4$\,ms results.

\section*{Theory}
\section{Stochastic Diffusion Model}\label{sec:stat. analysis}

We describe the spectral diffusion of the ZPL by a Wiener process, which can be modeled by the stochastic differential equation
\begin{align}
    \begin{aligned}
        \omega(t + \tau) &= \omega(t) + \sigma Z \sqrt{\tau} 
    \end{aligned}
    \label{eq:Wiener}
\end{align}
where $Z$ is a normally distributed random variable (white noise) $p(Z) = \mathcal{N}(0,1)$ with mean $\mu = 0$ and variance $\sigma_z = 1$. The fixed time step $\tau$ corresponds to the time between repeated line scans in our experiments.  
We use this model for studying the time dependent broadening of the spectral line as shown in Fig.~2b in the main text. For the data sets that we studied we did not find saturation of the inhomogeneous linewidth. Had the line stopped broadening, we would have done a more extensive analysis with the more appropriate Ornstein-Uhlenbeck process. The results that we obtained with the Ornstein-Uhlenbeck process (not shown) did not show better agreement with the experimental data. 

\begin{figure}[ht!]
    \center
    \includegraphics[width = .45\textwidth]{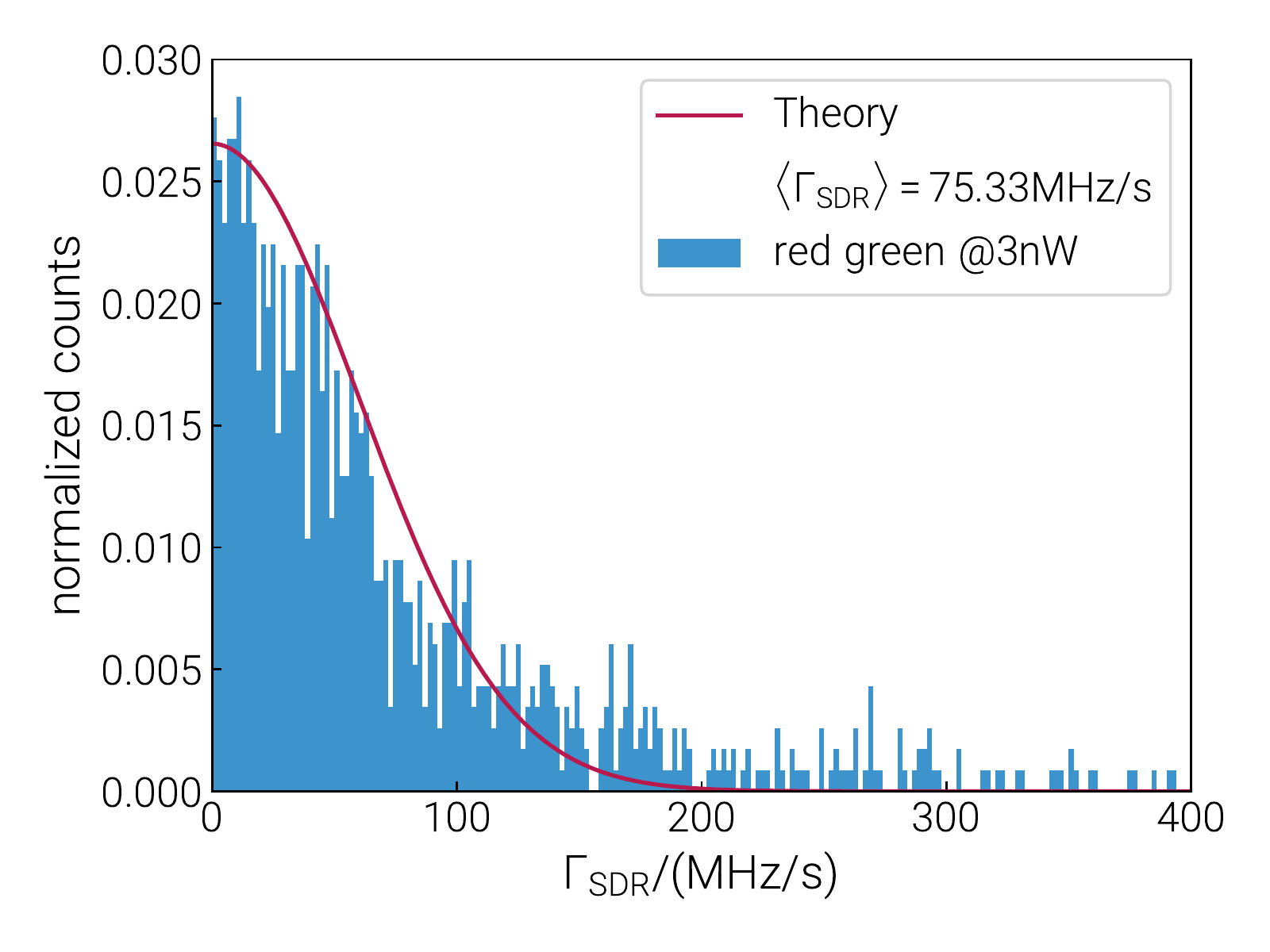}
    \caption{Spectral diffusion rate distribution. Probability distribution of $\Gamma_{\rm SDR}$ for red and green excitation light at $5{\rm nW}$. The distribution agrees well with the theoretical prediction (red) derived from the Wiener process at  $\eta(\omega_I, \tau) = 3.41 * 10^{24}/{\rm Joule/s^2}$ for a spectral diffusion rate of $75.35$\,MHz/s at $5$\,nW and $\tau = 2.3s $ which was used for modeling the diffusion process.}
    \label{Fig:sdr6nWredgreen}
\end{figure}

The spectral diffusion rate we investigated in our experiments is given by
\begin{align}
        \Gamma_{\rm SDR} = \frac{\langle |\omega(t+ \tau) - \omega(t)|\rangle}{\tau} = \frac{\sigma \langle |Z| \rangle}{\sqrt{\tau}}~, 
    \label{eq:SDRaverage}
\end{align}
which evaluates to 
\begin{align}
    \Gamma_{\rm SDR} = \sigma \sqrt{\frac{2}{ \pi \tau}}~. 
\end{align}
We observed that the the spectral diffusion rate scales like a square root power law with respect to the intensity. We can thus connect the spectral diffusion rate to an intensity by assuming 
\begin{equation}
    \sigma =  \sqrt{\eta(\omega_I, \tau) I} ~,
\end{equation}
where $\eta(\omega_I, \tau)$ is some proportionality constant that depends on the excitation frequency $\omega_I$ and the time step $\tau$. $I$ is the intensity of the excitation light. 
We thus find for the spectral diffusion rate
\begin{equation}
    \Gamma_{\rm SDR} = \sqrt{\frac{2 I \eta(\omega_I,\tau)}{\pi \tau}}~. 
\end{equation}
From the Eq.\eqref{eq:SDRaverage} it follows that $\Gamma_{\rm SDR}$ is the average of the absolute value of the random variable
\begin{equation}
    Z_{\Gamma_{\rm SDR}} = \frac{\omega(t+\tau) - \omega(t)}{\tau}  
\end{equation}
which by definition of the Wiener process \eqref{eq:Wiener} is normally distributed:
\begin{equation}
    p(Z_{\Gamma_{\rm SDR}}) = \mathcal{N}(0, \sigma /\sqrt{\tau })~.
\end{equation}
When we compare the simulation and the experimentally measured distributions of $Z_{\rm SDR}$ we find very good qualitative agreement (see for example Fig.\ref{Fig:sdr6nWredgreen}).
In Fig.~2b in the main text the results of using the the Wiener process to find the inhomogeneously broadened linewidth can be seen. The goal is to find a distribution of inhomogeneous linewidths for 14 trajectories, so that we can compare the simulated diffusion with the experimental results. For that purpose we numerically generate a large set of trajectories using Eq.\eqref{eq:Wiener} using spectral diffusion rate of $\Gamma_{\rm SDR} = 41$\,MHz/s and a time step of $0.8$\,s. The set of trajectories is then divided into smaller sets of 14 lines. We use the following method to determine the distribution of inhomogeneous linie widths: At a given time step the frequencies of 14 trajectories are taken as the center frequencies of 14 Lorentzians with a width of 60\,MHz. This spectrum is then fitted with a Voigt profile, whose FWHM determines the inhomogeneous linewidth. We repeat this for all trajectory subsets and record their respective FWHMs. For each time step we can then calculate the mean and the variance as shown in Fig.~2b in the main text. 

For smaller times the selected trajectories do not fall within one standard deviation of the simulated linewidths. This can either be attributed to the post selection, or subdiffusive behaviour. For larger times the mean of the simulated trajectories deviates significantly from the square root behaviour expected of a diffusion process. This can be attributed to using only very few spectral lines, which leads to large fluctuations in the fitting procedure. 
    
We also used an Ornstein-Uhlenbeck process to model the diffusion of the spectral line. We could not find a good qualitative agreement between the Ornstein-Uhlenbeck process and the broadening of the spectral line for various drift parameters. As discussed in a previous paragraph, the recorded trajectories did not have enough time to drift, to reproduce saturation of the inhomogeneous linewidth.

\section{Monte Carlo Simulation}\label{sec:Monte Carlo}

In this section we detail our analysis of the contribution of a fluctuating charge environment to the inhomogeneously broadened ZPL of the NV center. The NV's ZPL is sensitive to local electric fields due to the lack of inversion symmetry. A fluctuating charge environment produces a fluctuating local electric field at the position of the NV, which in turn Stark shits the position of the emission line. The accumulation of these shifts over time results in an inhomogeneously broadened spectral line. We investigated both the temporal change and the overall broadening in our experiments. The goal of this section is to make quantitative and qualitative estimates for the amount of broadening that is caused by a fixed number of charges in the NV's environment. Because we were able to observe trajectories of almost life time limited spectral lines we conclude that changes in the charge environment take place on timescales that are longer than the duration of individual line scans. We thus argue that a Monte Carlo simulation, in which we record the individual line shifts caused by a randomly distributing a fixed number of charges in the NV's environment is an adequate method to determine the line broadening produced by charge noise.  

To tailor the Monte Carlo simulation as best as possible to the experimental conditions we have to find the 
local electric field of a quasi static charge distribution in a diamond nano-pillar and determine the magnitude of the Stark shift. The tasks of finding the local field and its effect on the NV can be separated: In the following sections we present an expression for the electric field produced by the point charges inside a dielectric cylinder and a microscopic model of the interaction of a static field with the NV center. 
We then describe the details and the results of the Monte Carlo simulation. Charge screening is not systematically included in our simulations. However, screening charges would significantly dampen the electric field on length scales of the screening length and thus decrease the magnitude of field fluctuations at the location of the NV. This effect would lead to an overestimation of the SDR if the number of charges is kept fixed. When using the Thomas-Fermi model of screening we find a screening length of a few nano meter, which would eliminate most of the SD. This means that a more involved model of screening is needed, which, given the excitation scheme, accurately predicts the trapped charge density and the delocalized charge density. The delocalized charges will then produce the screening effect. Such a model is beyond the scope of this paper, but it would be highly valuable. It would clarify how screening could be used to mitigate SD, because screening reduces the impact of fluctuating charges on the NV's ZPL resonance.

\subsection*{Electric Field of a Charge in a dielectric cylinder}

The goal of this section is to present closed expressions for the electric field produced by a group of point charges in a diamond nano-pillar. This will allow us to efficiently determine the local electric field at the position of the NV. For this purpose we approximate the nano-pillar as a dielectric cylinder that has infinite length. 
This approximation will hold as long as the Stark shifts that are produced by surface charges on the tip of the cylinder are smaller than the natural linewidth. The extension of the pillar into an infinite cylinder allows us to use the results of \cite{Cui_2006}, where the electrostatic potential of a point charge in a cylindrical dielectric was calculated. According to \cite{Cui_2006} the potential of a point charge  (in cylindrical coordinates) in a dielectric cylinder is
\begin{align}
    \begin{aligned}
    \phi(\vec r,\vec r') = &\frac{q}{4 \pi^2 \epsilon_0 \epsilon_r }\int_0^\infty \d k \cos[k (z-z')]\times \left[ \frac{1}{2}I_0(k \rho_<) K_0(k \rho_>)+
    \sum_{m=1}^\infty I_m(k \rho_<) K_m(k \rho_>)\cos[m(\phi - \phi')] \right.\\
    &\left. -\frac{q_0(k) K_0(k R_0)}{2I_0(k R_0)}I_0(k \rho_<) I_0(k \rho_>)+
    \sum_{m=1}^\infty \frac{q_m(k)K_m(k R_0)}{I_m(k R_0)}\times I_m(k \rho_<)I_m(k \rho_>)\cos[m(\phi - \phi')]
    \right]
    \end{aligned}\label{eq:field_complete}
\end{align}
where $q$ is the charge, $I_m $ and $K_m$ are the modified Bessel functions, $R_0$ the pillar's radius, $\rho_<$ and $\rho_>$ are the smaller and bigger radii when comparing the observation point $\vec r$ and the source point $\vec r'$ and
\begin{equation}
q_m = \frac{1 - \epsilon''/\epsilon'}{1 - g_m(k) \epsilon'' /\epsilon'}
\end{equation}
where $\epsilon''$ ($\epsilon'$)is the relative permittivity inside(outside) the pillar and 
\begin{equation}
g_m = \frac{K_m(k R_0) I_m'(k R_0)}{K_m'(k R_0) I_m(k R_0)}
\end{equation}
The electric field is given by the gradient of the potential:
\begin{align}
\vec E(\vec r,\vec{r'})& = -\nabla \phi(\vec r, \vec r') \\ 
&= -(\pmb{\rho} \partial_\rho \phi(\vec r,\vec r') + \frac{1}{\rho} \pmb{\varphi} \partial_\varphi \phi(\vec r, \vec r') + \vec{z} \partial_z \phi(\vec r, r'))
\label{eq:nabla}
\end{align}
For the Monte Carlo simulation we assume that the NV is located on the symmetry axis of the cylinder. We can therefore place the coordinate origin at the location of the NV. 
The first two terms of the integrand in Eq.\eqref{eq:field_complete} do not converge quickly. Fortunately they evaluate to become (\cite{Cui_2006}) the ,,direct'' field of a charge, which is given by the potential:
\begin{align}
\begin{aligned}
\phi'(\vec r,\vec r') = &\frac{q}{4 \pi^2 \epsilon_0 \epsilon_r }\int_0^\infty \d k \cos[k (z-z')] 
\left[ \frac{1}{2}I_0(k \rho_<) K_0(k \rho_>)\sum_{m=1}^\infty I_m(k \rho_<) K_m(k \rho_>)\cos[m(\phi - \phi')] \right]\\
&=\frac{q}{4 \pi\epsilon_0 \epsilon_r }\frac{1}{\vec r - \vec r'}
\end{aligned}
\end{align}
If the observation point ($\vec r$) lies on the cylinder axis the derivation with respect to $\phi$ vanishes. The electric field at the origin thus becomes:
\begin{align}
\begin{aligned}
\tilde{\vec{E}}(\vec 0, \vec r') = -&\frac{q}{4 \pi^2 \epsilon_0 \epsilon_r }\left( \vec{z} \partial_z + \bm{\rho} \partial_{\rho} \right)\int_0^\infty \d k  cos[k(z-z')] \frac{q_0(k) K_0(k R_0)}{2I_0(k R_0)}I_0(k \rho_<)  \left.I_0(k \rho_>)\right|_{\vec r = \vec 0}
\end{aligned} ~,\label{eq:integral-field}
\end{align}
where $\vec{z}$ and $\bm{\rho}$ are unit vectors in $z$ and $\rho$ direction
so that 
\begin{align}
\vec{E}(\vec 0, \vec{r'}) = -\nabla \phi'(\vec r, \vec r')|_{\vec{r = 0}} + \tilde{\vec{E}} (\vec{0},\vec r')~.
\label{eq:pol-correction}
\end{align}
The second term in the above equation can be interpreted as the effect of the surface polarization charge.
For a group of charges the field generated by multiple charges at the origin is given by
\begin{equation}
\vec E = \sum_n \vec{E}(\vec 0, \vec r_n)\label{eq:total-field}
\end{equation}
where $\vec{r}_n$ is the position of the n'th charge. The integral in Eq.\eqref{eq:integral-field} is calculated numerically. 

\subsection*{Interaction of NV with electric field}

According to \cite{Maze_2011} the coupling of the electric field to the excited states is given by 
\begin{equation}
H_E = g(b+d)E_z + ga M(\vec{E}) \label{eq:ex-coupling}
\end{equation}
where 
\begin{equation}
M(\vec{E}) = \begin{pmatrix}
\vec 0 & \vec 0 & E_x - E_y \sigma_y  \\
\vec 0 & E_x \sigma_z + E_y \sigma_x & \vec 0 \\
E_x - E_y \sigma_y & \vec 0 & \vec 0 
\end{pmatrix}~,
\end{equation}
$\vec{E} = (E_x, E_y, E_z)$, $\sigma_i$ are the Pauli matrices and $\vec 0$ is a two by two matrix with all entries equal to zero. Eq.\eqref{eq:ex-coupling} is stated in the basis $\{ A_1, A_2, \tilde{E}_x, \tilde{E}_y, E_1, E_2\}$. We labeled the $\tilde{E}_x, \tilde{E}_y,$ orbitals with a tilde, to avoid confusion with the respective electric field components. The ground state tripled couples to the electric field according to 
\begin{equation}
H_E = 2 g b E_z
\end{equation}
in the Basis $\{ {}^3A_{2+}, {}^3A_{20},{}^3A_{2-}\}$. 
The coupling parameters are given by
\begin{align}
& g \approx 2 {\rm pHz}\\
& a \approx b \approx c \approx 0.3 \mu ({\rm MV/m})^{-1}\\
& d \approx 3 \mu ({\rm MV/m})^{-1}
\end{align}
To compare the Monte Carlo simulation with the experiment we choose the $\tilde{E}_{x,y} \rightarrow {}^3A_{20}$ transition to determine the change of the transition energy due to the presence of an electric field. For this purpose we diagonalize the above Hamiltonians for a given electric field and record the relative shifts in energy compared to the unperturbed energy of the $\tilde{E}_{x,y} \rightarrow {}^3A_{20}$ transition. The Stark shifts are given by
\begin{align}
   \Delta_{\pm} = (b - d) g E_z \pm a g\sqrt{E_x^2 + E_y^2} ~. \label{eq:stark-shift}
\end{align}
The electric field lifts the degeneracy of the $\tilde{E}_{x,y}$ orbitals. In the experiments we only resolve only one spectral line. For the Monte Carlo simulations we only use $\Delta_-$, because we assume that the other transition is not bright enough to be observed. 

\subsection*{Results} \label{sec: MC results}

\subsubsection*{Bulk}

\begin{figure}[ht!]
    \centering
    \includegraphics[height = 5cm]{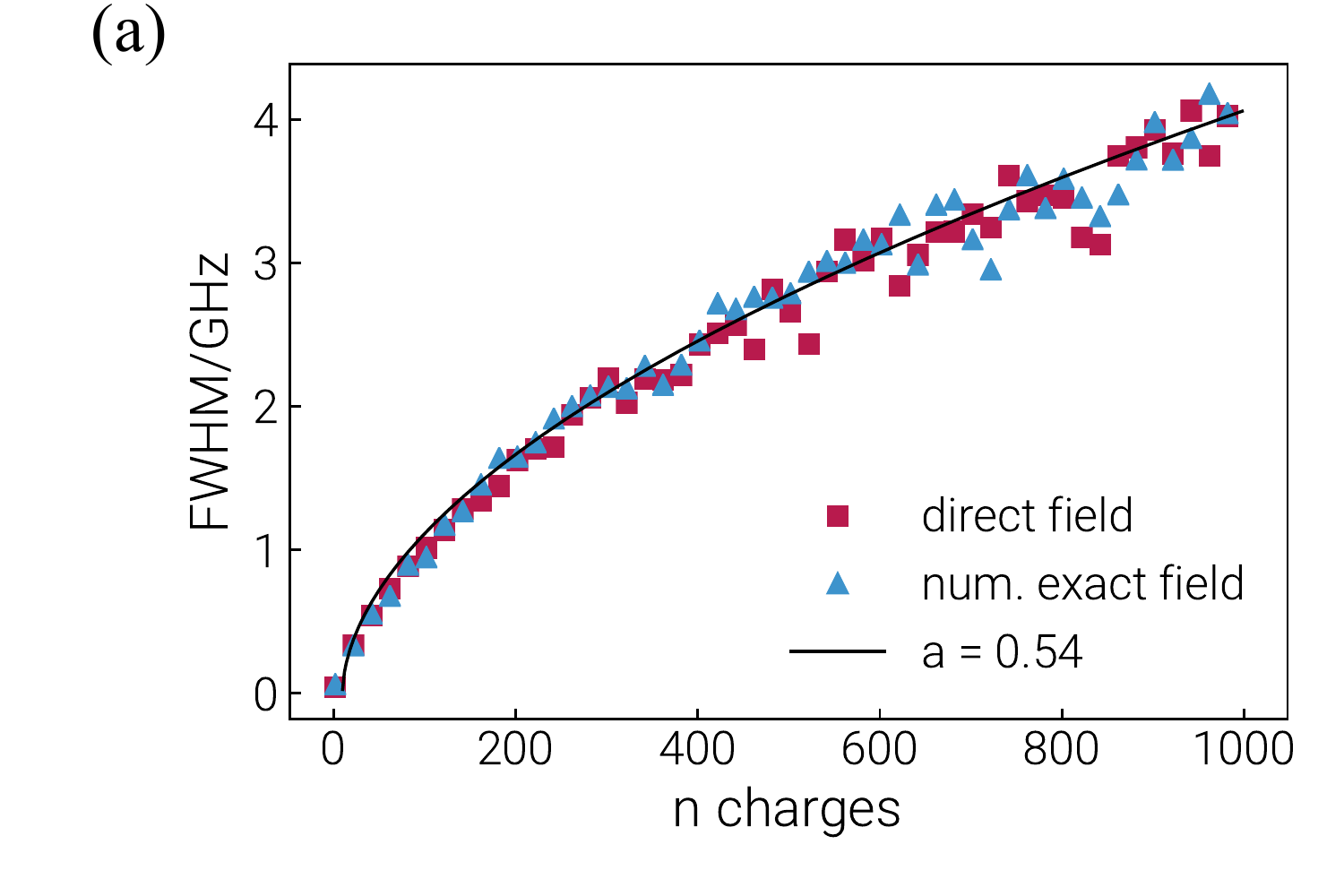}
    \includegraphics[height = 5cm]{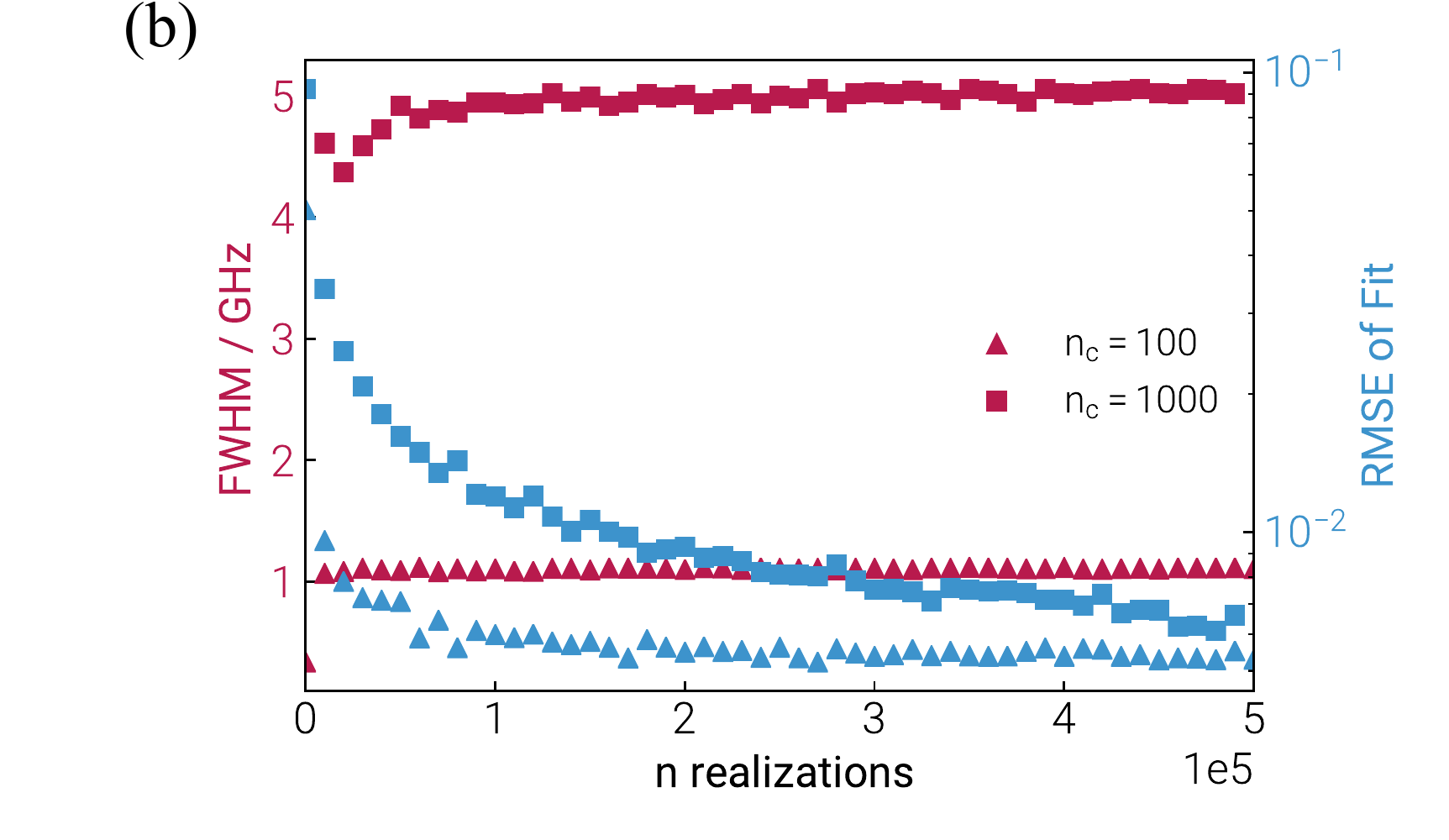}
    \caption{ Inhomogeneous broadening. (a) Inhomogeneous broadening caused by random distribution of charges with a density of $1$ ppm (approx. $13800$) in a cylindrical volume with a height of $1600$\,nm and a radius of $125$\,nm. 
    The blue triangles are calculated using the electric field including the contribution produces by the polarization surface charge [$\tilde{\vec{E}}$ in Eq.\eqref{eq:total-field}]. The red squares show the result of only including the direct field. The black line represents a fit with the powerlaw $f(x) = b (x- x_0) ^ {a}$. (b) The blue triangles (100 charges) and squares (1000 charges) mark the FWHM of the inhomogeneous line for two different charge densities depending on the number of realizations of distributing the charges among the charge traps. The red squares and triangles show the root means square error (RSME) of the fit and the spectrum. The error drops for a higher count of realizations and appears to be saturating, indicating, that the resulting spectrum reaches its equilibrium value. We chose $1000$ realizations for the Monte Carlo simulation in (a).}
\label{Fig:bulk}
\end{figure}

In this section we present the results of the Monte Carlo simulation of the inhomogeneous linewidth produced by electric charges in the environment of the NV. For now we do not consider any charge traps on the surface of the pillar and only focus on charges in the bulk. For the simulation we randomly distribute charge traps in in a cylindrical volume with the dimensions of the pillars used in the experiment ($r = 125 {\rm nm}$, $h = 1600 {\rm nm}$). We set the density of the charge traps to 1 ppm (approx. $13800$ traps) to imitate the density of the P1 centers, which we assume are the primary source of electric charge noise in our problem. P1 centers are likely to be electron donors which makes them a likely candidate for a stationary charge impurity that can change its charge state during repeated line scans. 
A single iteration of our simulation consists of randomly distributing a fixed number of charges among the charge traps. The local electric field produced by the charges, which is calculated according to Eq.\eqref{eq:total-field} causes a change in the position of the resonance frequency as described by Eq.\eqref{eq:stark-shift}. We assume global charge neutrality ($\sum_n q_n = 0$). After a fixed set of iterations the frequency shifts are taken as the center frequencies of lifetime limited Lorentzians. The resulting spectrum is the sum of all the Lorentzians. We fit the result with a Voigt profile to determine the linewidth of the inhomogeneously broadened line. The result of the procedure can be seen in Fig.~\ref{Fig:bulk}a. For the simulation $10000$ random charge distributions were sampled to produce a point in Fig.\ref{Fig:bulk} a. The inhomogeneous linewidth showed reasonable saturation for $10000$ iterations [Fig.~\ref{Fig:bulk}b]. In Fig.~\ref{Fig:bulk}a, no charge traps were placed on the surface of the cylinder. The surface charges are considered separately in the next paragraph. The correction caused by surface polarization charges [$\tilde{\vec{E}}$ in Eq.\eqref{eq:total-field}] is almost negligible for the chosen geometry and the positioning of the NV center. As can be seen by comparing the squares and triangles dots in Fig.~\ref{Fig:bulk}a.     

\subsubsection*{Surface}

\begin{figure}[ht!]
    \center
    \includegraphics[width = .45 \columnwidth]{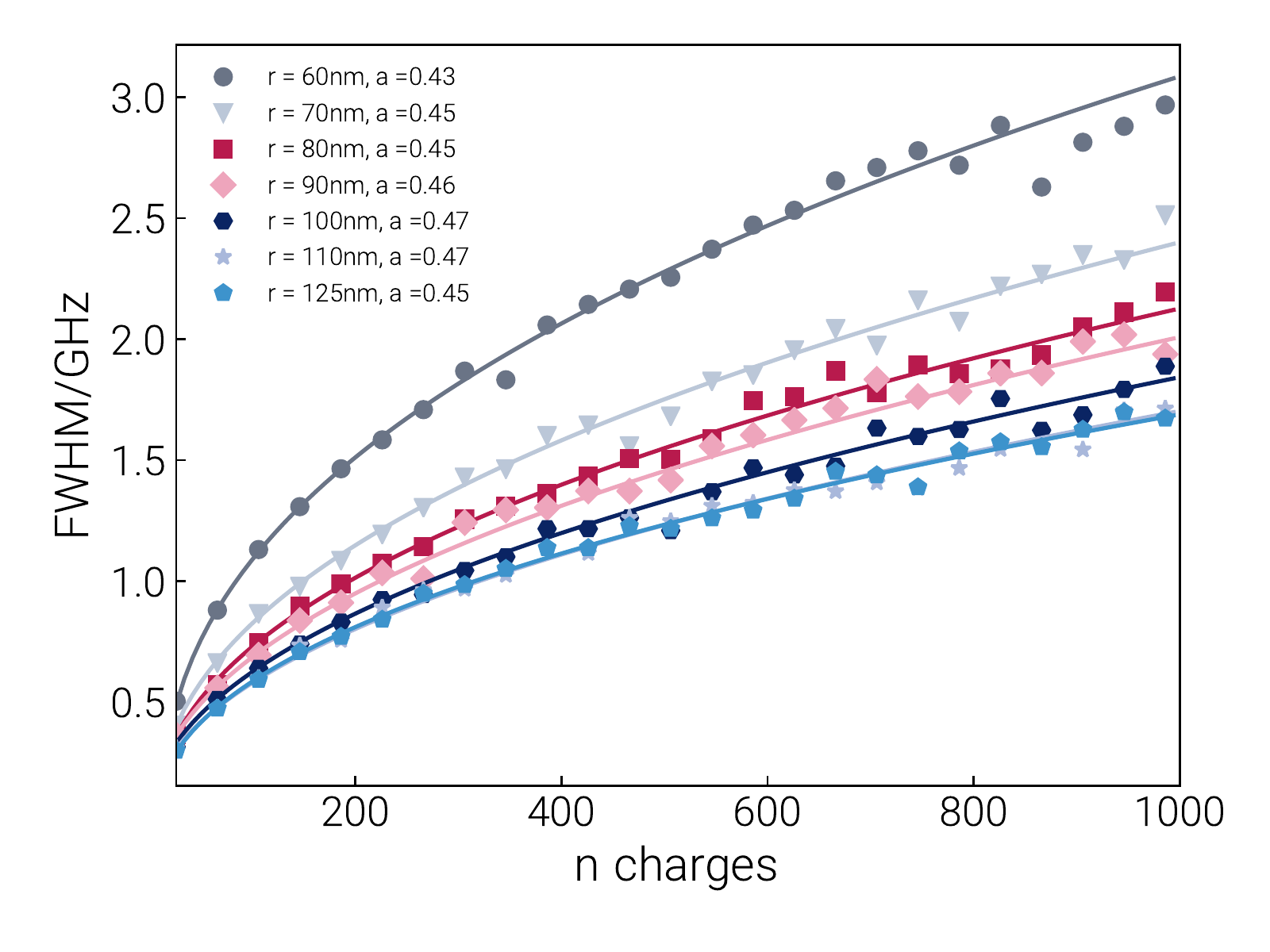}
    \caption{ Effect of surface charges on inhomogeneous linewidth. Inhomogeneous broadening due to surface charges for different pillar raddii. The lines represent a fit with the powerlaw $f(x) = b (x- x_0) ^ {a}$. The power laws noticeably differ from the power law found for bulk charges [Fig.~\ref{Fig:bulk}a].}
    \label{Fig:surface_var_r}
\end{figure}

Here we repeat the previous analysis but we focus on surface defects. We are applying the same reasoning as in the previous section, but are limiting trap locations to the cylinder surface ($6000$ in total). This way we can make an estimate on the potential impact of surface charges and if they are contributing to the field noise at the location of the NV. The impact of the trap density on the linewidth is no longer a big concern for larger densities. This can be seen for example in Fig.~\ref{Fig:trap-impact} where we investigated the inhomogeneous linewidth for different trap densities and a fixed amount of charge. The surface can have a significant impact on the broadening as can be seen in Fig.~\ref{Fig:surface_var_r}, where the inhomogeneous broadening is shown for a range of different pillar radii. The impact of surface charges on the line broadening is more and more noticeable for smaller radii. It is interesting to observe that the power law fits in Fig.~\ref{Fig:surface_var_r} give exponents that are consistently below $0.5$. This distinguishes a pure surface charge contribution to the broadening from randomly placing charges into the cylinder volume.     

\subsubsection*{Surface vs. Bulk}

\begin{figure}[ht!]
    \center
    \includegraphics[width = .45 \columnwidth]{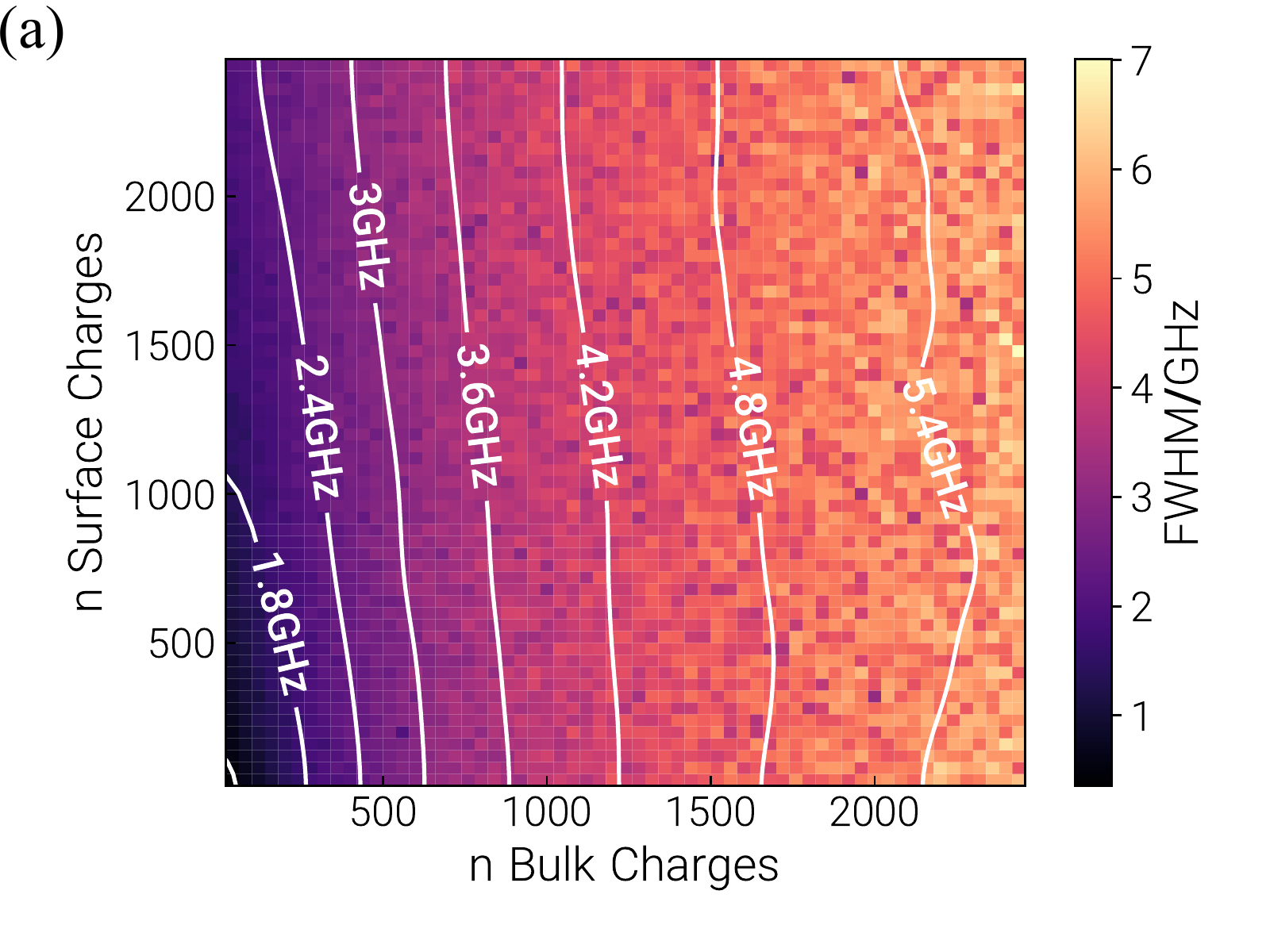}
    \includegraphics[width = .45 \columnwidth]{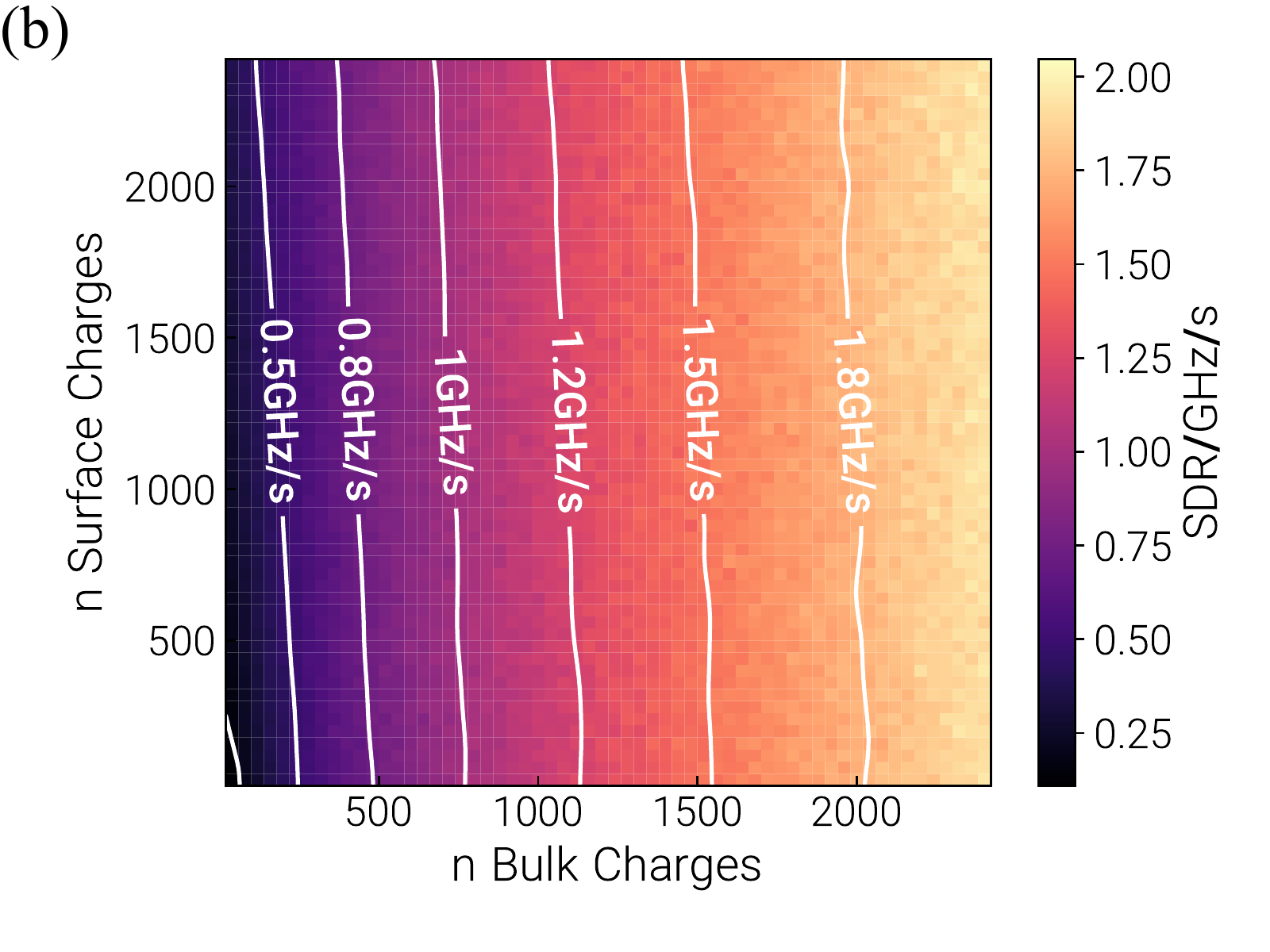}
    \caption{ Joint effect of surface and bulk charges. (a) Inhomogeneous broadening for a range of surface and bulk charges.  (b) SDR calculated by fixing the time step per reordered charge configuration at 2000 bulk charges, which corresponds to a broadening of approximately $5{\rm GHz}$ (without screening). }
\label{Fig:surface_v_bulk}
\end{figure}

\begin{figure}[ht!]
    \center
    \includegraphics[width = .45 \columnwidth]{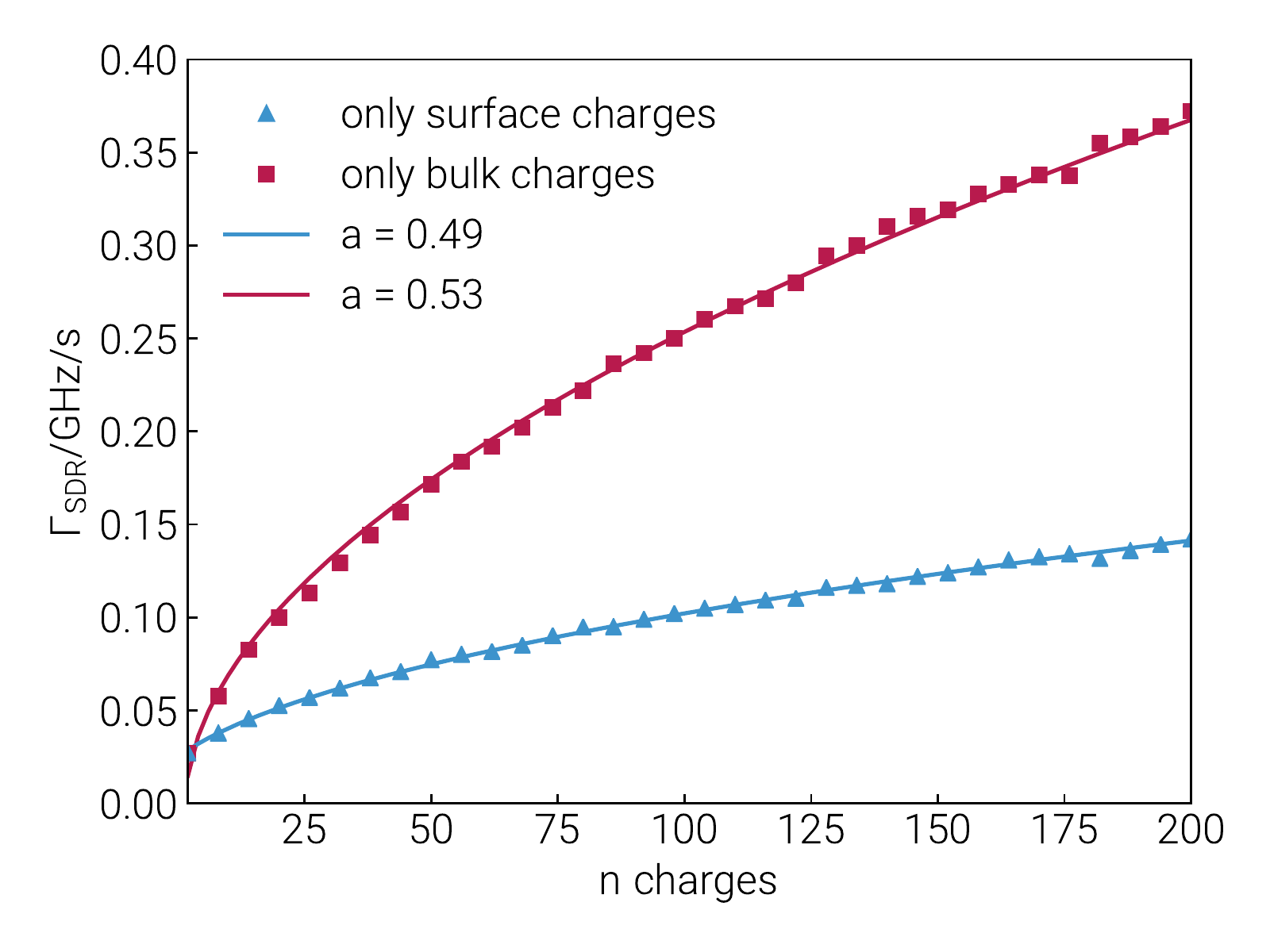}
    \caption{ Spectral diffusion rate scaling behaviour. Shown are the SDR for no bulk charges (red triangles) and no surface charges (green squares) as a function of the number of charges contributing to the diffusion process. The parameters are the same as for Fig.~\ref{Fig:surface_v_bulk}. The solid lines correspond to a power law fit with $\Gamma_{\rm SDR} = b (n - n_0)^a$ .The power laws show a surprising similarity to the power laws in Fig.~3c in the main text.}
\label{Fig:surface_v_bulk_SDR_comp}
\end{figure}

In this section we discuss the results of the impact on the line broadening of both, fluctuating surface and bulk charges.
Again we assume a pillar with $h = 1600{\rm nm}$, $r = 125 {\rm nm}$ and a NV that is located at its center. In each run of the simulation a fixed amount of surface and bulk charges are allowed to assume random positions determined by fixed trap locations. 
The spectral diffusion rate is given by the difference of the Stark shifts of two subsequent charge configurations divided by a time step. Because the Monte Carlo simulation lacks any real time, an ad hoc time step has to be introduced such that the spectral diffusion rate and the broadening as predicted by the simulation match the experimental data for a fixed intensity. Here we use the spectral diffusion rate of 1730\,MHz/s and a total inhomogeneous broadening of 5\,GHz. The inhomogeneous linewidth was measured by integrating over thousands of PLE scans at 37\,nW excitation power and by applying strong green pulses between the PLE scans. The simulation method could in principle allow us to predict the average amount of charge that contributes to the to diffusion process by measuring the spectral diffusion rate. The results of the simulations can be seen in Fig.~\ref{Fig:surface_v_bulk}a and b . We refrain from inferring real world charge densities from comparing the Monte Carlo simulations with the experimental data. For a satisfactory comparison charge screening has to be included in the simulation. Without screening the Monte Carlo simulation underestimates the amount of charge contributing to the diffusion process. We also could not determine the ad hoc time step for all excitation schemes, because the linewidth did not always saturate (see for example Fig.~2c in the main text). The qualitative behavior is however not significantly impeded by these shortcomings of the simulation and we see similarities in the scaling behaviour of the SDR in the simulation and the experiment (compare Fig.~3c in the main text and Fig.~\ref{Fig:surface_v_bulk_SDR_comp}) 

\subsubsection*{Trap densities}

\begin{figure}[ht!]
    \center
    \includegraphics[width = .45 \textwidth]{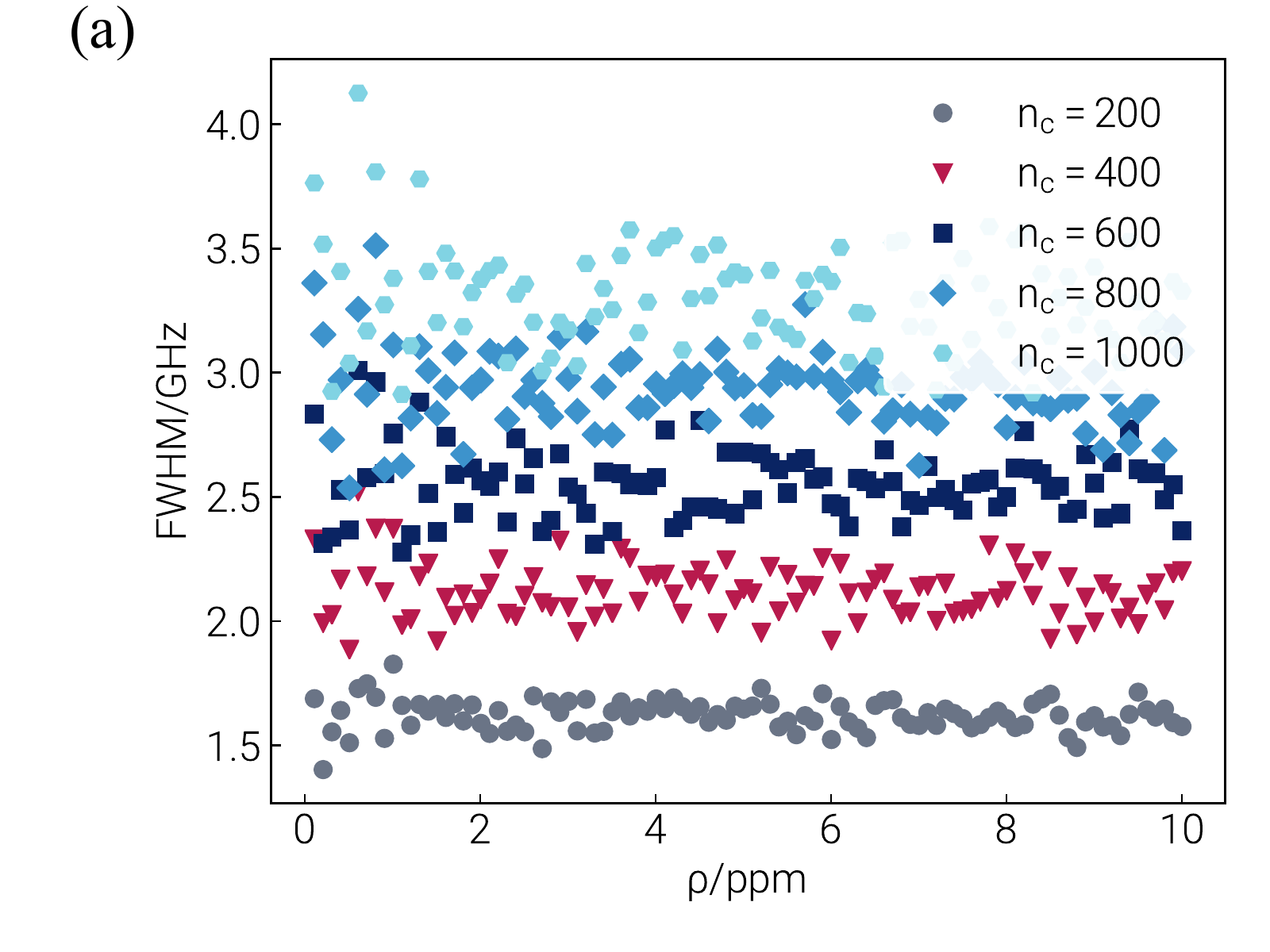}
    \includegraphics[width = .45 \textwidth]{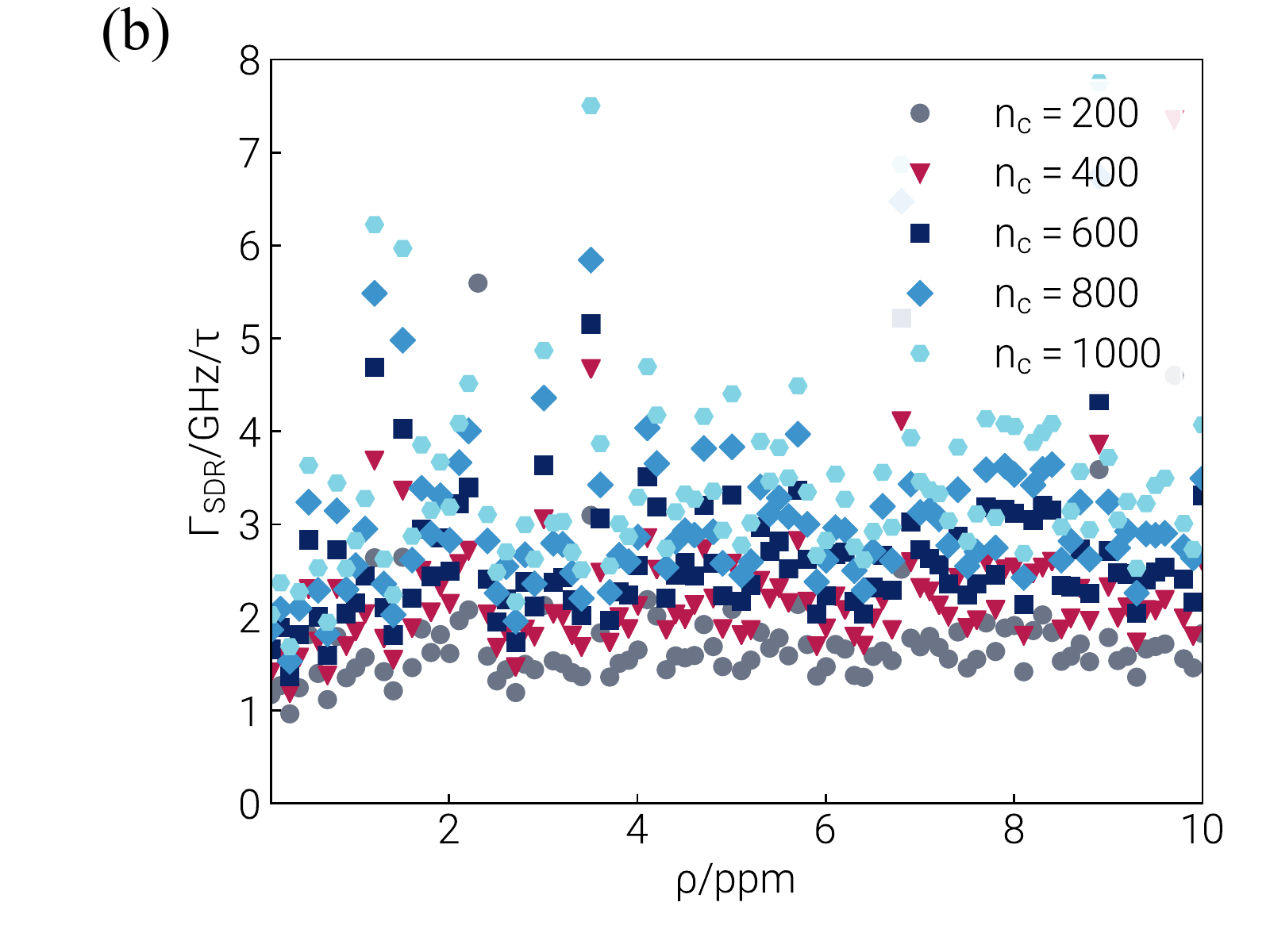}
    \caption{ Effect of trap densities. Shown the inhomogeneous linewidth (a) and the spectral diffusion rate (b) as a function of the trap density $\rho/{\rm ppm}$ for a range of charges and for 10000 different configurations. The trap density appears to have little impact on the line broadening and the spectral diffusion range. The time step for the spectral diffusion rate is fixed to some arbitrary value, as it does not impact the qualitative behavior of the spectral diffusion rate.}
\label{Fig:trap-impact}
\end{figure}

In this section we briefly comment on the relation between trap densities and the inhomogeneous line broadening. Performing the Monte Carlo simulations for different trap and charge densities revealed a negligible impact of the trap densities on the inhomogeneous linewidth. The same appears to hold for the spectral diffusion rate, which is given by the energy difference of two subsequent charge configurations divided by some arbitrary time step. The time step is not important here, as it does not impact the results. In Fig.~\ref{Fig:trap-impact} we show both the impact of changing the trap density on the line broadening and the spectral diffusion rate. The result is somewhat surprising, as trap  densities $\rho> 2 \rm ppm$ do not significantly decrease the spectral diffusion. The amount of charge that is free to fluctuate has a much greater impact for larger densities. A decrease in the spectral diffusion rate can be seen for densities $\rho <2 \rm ppm$. Here a noticeable reduction in the rate of spectral jumps can be seen for smaller charge densities (see e.g. the blue circles in Fig.~\ref{Fig:trap-impact}b $n_c = 200$). 

\bibliography{bibliography}